\documentclass[final,3p,times,twocolumn,numbers,sort&compress]{elsarticle}

\usepackage{graphicx}
\usepackage{amssymb}
\usepackage{lineno} 
\usepackage{hyperref}
\usepackage{comment}
\usepackage{color,soul}

\journal{Nucl. Instrum. Methods Phys. Res., A}

\begin{document}

\begin{frontmatter}

\title{Atmospheric ray tomography for low-Z materials:\\implementing new methods on a proof-of-concept tomograph}



\author[ICV,PWC]{Gholamreza Anbarjafari}
\author[GScan]{Aivo Anier}
\author[GScan,ICV]{Egils Avots}
\author[NANU,GScan,UT]{Anzori Georgadze}
\author[GScan,KBFI]{Andi Hektor\corref{cor1}}
\ead{andi.hektor@cern.ch}
\author[GScan,UT]{Madis Kiisk}
\author[GScan]{Marius Kutateladze}
\author[GScan,UT]{Tõnu Lepp}
\author[GScan]{Märt Mägi}
\author[GScan,UT]{Vitali Pastsuk}
\author[GScan]{Hannes Plinte}
\author[UT]{Sander Suurpere}
\address[GScan]{GScan OU, Maealuse 2/1, 12618 Tallinn, Estonia}
\address[ICV]{iCV Lab, University of Tartu, Narva mnt 18, Tartu 51009, Estonia}
\address[UT]{Institute of Physics, University of Tartu, W. Ostwaldi tn 1, 50411 Tartu, Estonia}
\address[KBFI]{KBFI, Akadeemia tee 23, 12618 Tallinn, Estonia}
\address[NANU]{Institute for Nuclear Research NANU, 03028 Kyiv, Ukraine}
\address[PWC]{PwC Advisory, Helsinki, Finland}
\cortext[cor1]{Corresponding author.}

\begin{abstract}
Cosmic rays interacting with the atmosphere result in a flux of secondary particles including muons and electrons. Atmospheric ray tomography (ART) uses the muons and electrons for detecting objects and their composition. This paper presents new methods and a proof-of-concept tomography system developed for the ART of low-Z materials. We introduce the Particle Track Filtering (PTF) and Multi-Modality Tomographic Reconstruction (MMTR) methods. Based on Geant4 models we optimized the tomography system, the parameters of PTF and MMTR. Based on plastic scintillating fiber arrays we achieved the spatial resolution 120~$\mu$m and 1~mrad angular resolution in the track reconstruction. We developed a novel edge detection method to separate the logical volumes of scanned object. We show its effectiveness on single (e.g. water, aluminum) and double material (e.g. explosive RDX in flesh) objects. The tabletop tomograph we built showed excellent agreement between simulations and measurements. We are able to increase the discriminating power of ART on low-Z materials significantly. This work opens up new routes for the commercialization of ART tomography.
\end{abstract}

\end{frontmatter}


\section{Introduction}
\label{Sec:1}

Coulomb scattering based muon tomography was proposed in 2003~\cite{Borozdin2003aa}. Since then many research groups have demonstrated that the method can be successfully implemented in different fields to detect high-Z materials, e.g., nuclear safety~\cite{Clarkson:2014xfa, Poulson:2016fre, Mahon:2019aa} or security applications~\cite{Morris:2008aaa, Riggi:2018aaa, Harel:2019aaa}. In recent years several reviews have appeared providing excellent insights into last developments~\cite{Bonechi:2019ckl, Kaiser:2019aa, Vanini:2019aab, Procureur:2018aaa, Bonomi:2017uow}.

Coulomb scattering depends strongly on atomic number $Z$. It makes the recognition of high-Z materials (e.g. nuclear materials) in muon tomography or more generally, in atmospheric ray tomography (ART), easier than the recognition of low-Z materials. Thus ART applications on low-Z materials have merged slowly. Technological challenges are particularly complex in security applications, where low-Z materials should be identified within minute-scale time.

However, some dedicated research papers have appeared on the feasibility of the low-Z ART for security applications. Klimenko et al have studied whether electrons can be used for flux attenuation as an additional signature in muon tomography~\cite{Klimenko:2005aaa}. Other accompanying physical effects such as muonic X-rays and electromagnetic showers have also been investigated. The authors summarize that the inclusion of the latter requires additional detection systems. Cuellar et al reported results based on Geant4 simulations that the stopping effect of the low momentum part of the leptons spectrum could be used as a source of additional information to scattering tomography, especially for medium and low-Z materials~\cite{Cuellar:2009aa}. Later the effect has been demonstrated with simulated and experimental results~\cite{Blanpied:2015aaa}. The authors have proposed to exploit the ratio of stopping power to scattering, which enables to eliminate the sample thickness as an unknown variable. First, one identifies the material from the ratio, then the mean scattering angle and the known radiation length can be computed to calculate the thickness. The simulated results with materials of 1~m$^2$ for heavy metal plates or cubes of 1~m$^3$ for paper and nylon, as well as experimental results on the pallet-size units with low-Z materials, were presented with 30, 10 and 5 minute long measurements.

The detection of explosives and narcotics has been investigated at the TUMUTY facility based on the Geant4 simulations, where the material classification was done using the machine learning (ML), Support Vector Machine (SVM)~\cite{Yifan:2018aac}. The authors conclude that ART allows to discern narcotics and explosives from background and metals, but not be separated from each other (the various $20\times20\times20$~cm$^3$ size objects with 10 to 30~minute measuring time were studied). The experimental validations of the numeric results on an RPC-based tomography system concluded that the actual spatial resolution (0.6~mm) was lower than previously were simulated~\cite{Pan:2019aaa}. Four materials were tested: flour (substitute of narcotics), aluminium, steel, and lead. The overall discrimination rate with the SVM classifier could reach 70, 95, and 99\% with 1, 5, and 10 minutes measurement, accordingly.

We have found only a few examples of low-Z ART in non-security applications. Bikit et al demonstrated that low-Z materials can be imaged by the simultaneous detection of cosmic muons and photons created via bremsstrahlung inside the investigated bone and soft tissue sample~\cite{Bikit:2016abc, Mrdja:2016aaa}. Another example is inspection of nuclear waste encapsulated in concrete matrix. The investigations on the containers of nuclear waste showed that litre-size gas bubbles in concrete due to uranium dioxide oxidation can be discovered within days of exposure~\cite{Dobrowolska:2018abc}. Two recently papers reported that ART can detect reinforcement rebars in concrete. The first study is based on Monte Carlo simulations~\cite{Dobrowolska:2020aaa} and the second on experimental tests~\cite{Niederleithinger:2020wrf}.

Three main factors influence the ART performance: (i) the accuracy of particle trajectory estimation, (ii) ability to determine particle type and energy, (iii) the tomographic reconstruction methods deployed. The fourth, a system independent factor, is the measurement time available. In the early days of muon tomography, a geometry-based reconstruction approach, Point of Closest Approach (PoCa)~\cite{SCHULTZ2004687}, was introduced. PoCa is approximating multiple scattering processes with a single interaction point. It shows fairly good results in detection of high-Z materials submerged to low-Z materials. In conventional X-ray Computed Tomography (CT) analytical methods such as filtered back projection (FPB) have been historically widely used for its computational lightness. In CT the positions of source and detector are well known, but non-isotropic exposure and high Poisson noise in ART reduce its performance considerably. Iterative methods seem to be most suited, either algebraic~\cite{Liu:2019aab} or statistical approaches~\cite{Schultz:2007aaa, PESENTE:2009aaa}, where the best trajectory estimates are obtained using scattering and displacement data.

Due to the properties of the Coulomb multiple scattering, the performance of tomographic reconstruction is greatly improved if the energy of passing particle is estimated. For example, the measured angular distribution of passing particles allows to classify the particles according their estimated energy~\cite{Morris:2012aca}. The proposed momentum multi-group model method was experimentally tested and the reconstruction performance was quantitatively compared against the squared average, median angle squared and the average of squared angle using the receiver operating characteristic (ROC) curves~\cite{Perry:2014bad}. The results demonstrate clear superiority for multi-group approach. The CRIPT-project presented an another approach of momentum estimation employing massive iron sheets between their tracking detectors~\cite{ANGHEL:2015abs}.

The accurate reconstruction of particle tracks is especially important in low-Z ART and in applications having sub-cubic-meter volumes scanned. The resolution of scattering angle at mrad is required. The scattering density of muons is directly related to the atomic number $Z$ of material. The density is an order of magnitude lower for materials with low-Z (Ca or lighter elements, $Z \lesssim 20$) in comparison to high-Z (Pb or heavier elements, $Z \gtrsim 80$). For example, many experiments have applied the cylindrical drift chamber detectors in low-Z ART. The spatial resolution of small scale drift tubes can go down to 400~$\mu$m (FWHM) and angles of 2~mrad (FWHM)~\cite{Morris:2008aaa}. For large scale systems, a sub-mm position resolution perpendicular to the drift tube wire has been reported~\cite{Blanpied:2015aaa}. For an another type, resistive plate chambers (RPC), spatial resolution of 0.6~mm on lab-scale has been achieved~\cite{Pan:2019aaa}. The authors show no data on the spacing between the detector plates, so we cannot estimate the angular resolution.

On the sea level, the electrons and positrons constitute about 30-40\% of the total lepton flux and about a half of that is in practical usable energy range~\cite{Klimenko:2005aaa}. Due to low energy, this fraction of the lepton spectrum is strongly influenced from the environment around the detector system, for example, the building construction structures above the detector system. However, one can employ it as an additional source of information for the detection of small and medium sized objects, such as passenger luggage or similar scale objects.
 
In this paper we present new methods and a proof-of-concept tomography system developed specifically for the ART of low-Z materials. We introduce the Particle Track Filtering (PTF) and Multi-Modality Tomographic Reconstruction (MMTR) methods. PTF allows to classify the passing particle events (of charged leptons) into different groups so that the absorption and the scattering effect are more isolated. Some aspects of the PTF method are described in a patent application by some authors of the paper~\cite{Patent:2018aaa}.

In tomographic reconstruction we have developed a novel edge detection method to separate the logical volumes of scanned object. We demonstrated the effectiveness of the method in case of single composition material (e.g. water, aluminum, steel) and double composition material (e.g. explosive RDX in flesh-like material) objects. The MMTR approach enables to combine multiple reconstructions obtained from the different PTF-classified particle event groups and the scattering and transmission parameters from the track reconstruction.

We have composed the numerical models using the Geant4 software package~\cite{Allison:2006ve} to optimize the tomography system and the parameters of PTF and MMTR procedures. For the tracker system (the hodoscope) we chose plastic scintillating fibers arrays. With this tracker we were able to achieve a spatial resolution of 120~$\mu$m for muons and the 1~mrad angular resolution in the particle track reconstruction.

To validate our theoretical and numerical tests we built a physical proof-of-concept tomography system. The measurements on the prototype showed excellent agreement with the numerical simulations. We are able to increases significantly the discriminating power of ART for low-Z materials.

The structure of the paper is the following. We start with the description of PTF realized on plastic scintillator fiber arrays. In Section~\ref{Sec:3} we present the MMTR method. In both sections we validate the methods with numerical Geant4 models. In Section~\ref{Sec:4} we describe the general design principles of the physical proof-of-concept ART system. In Section~\ref{Sec:5} we give the technical description of the prototype and the first results of the measurements. In the last section we summarize the results and give some hints on the next steps we have considered.

\section{The particle track classification: the PTF method}
\label{Sec:2}

\begin{figure}[ht]
\centering\includegraphics[width=1.0\linewidth]{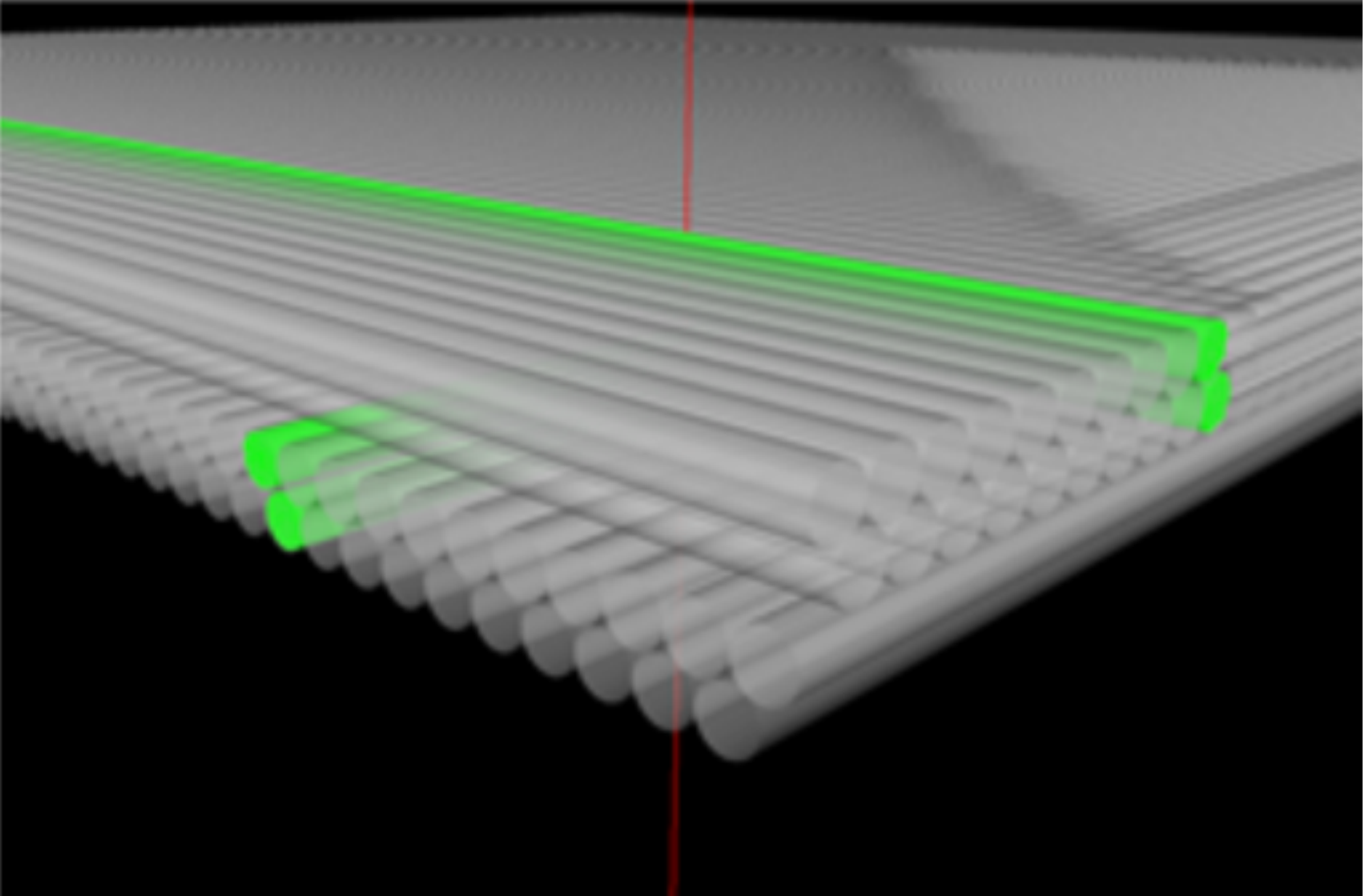}
\caption{A snapshot from the Geant4 model of the position-sensitive detector plate composed by the two orthogonally placed double-fiber-layers (gray cylinders). The red line denotes the trajectory of a passing muon. The greened fibers illustrate the propagation of scintillation light generated in the fibers by the muon passage.}\label{Fig:1}
\end{figure}

In this section we describe the main elements of the Particle Track Filtering (PTF) method. We have developed a virtual hodoscope system in Geant4 addressing the critical factors of the low-Z ART described above. Mainly, we address the accuracy of particle track reconstruction and the ability to classify the type and energy of particle event candidates. The hodoscope consists of the three position sensitive detector plates composed by plastic scintillator fibers. The detector plate has 4-layered structure: the two orthogonally placed double-layered fiber-mats, which give the $x$ and $y$ position of the particle hit (see Fig.~\ref{Fig:1}). The fiber diameter is 1.0~mm, the pitch in a single layer is 1.1~mm. The top layer is shifted by a half pitch, its position has been aligned according to the positions of the lower layer fibers. This ensures close to 100\% geometrical detection efficiency at the every angle of an incidence and high spatial resolution (for details, see Sec.~\ref{Sec:4}).

\begin{figure}[ht]
\centering\includegraphics[width=1.0\linewidth]{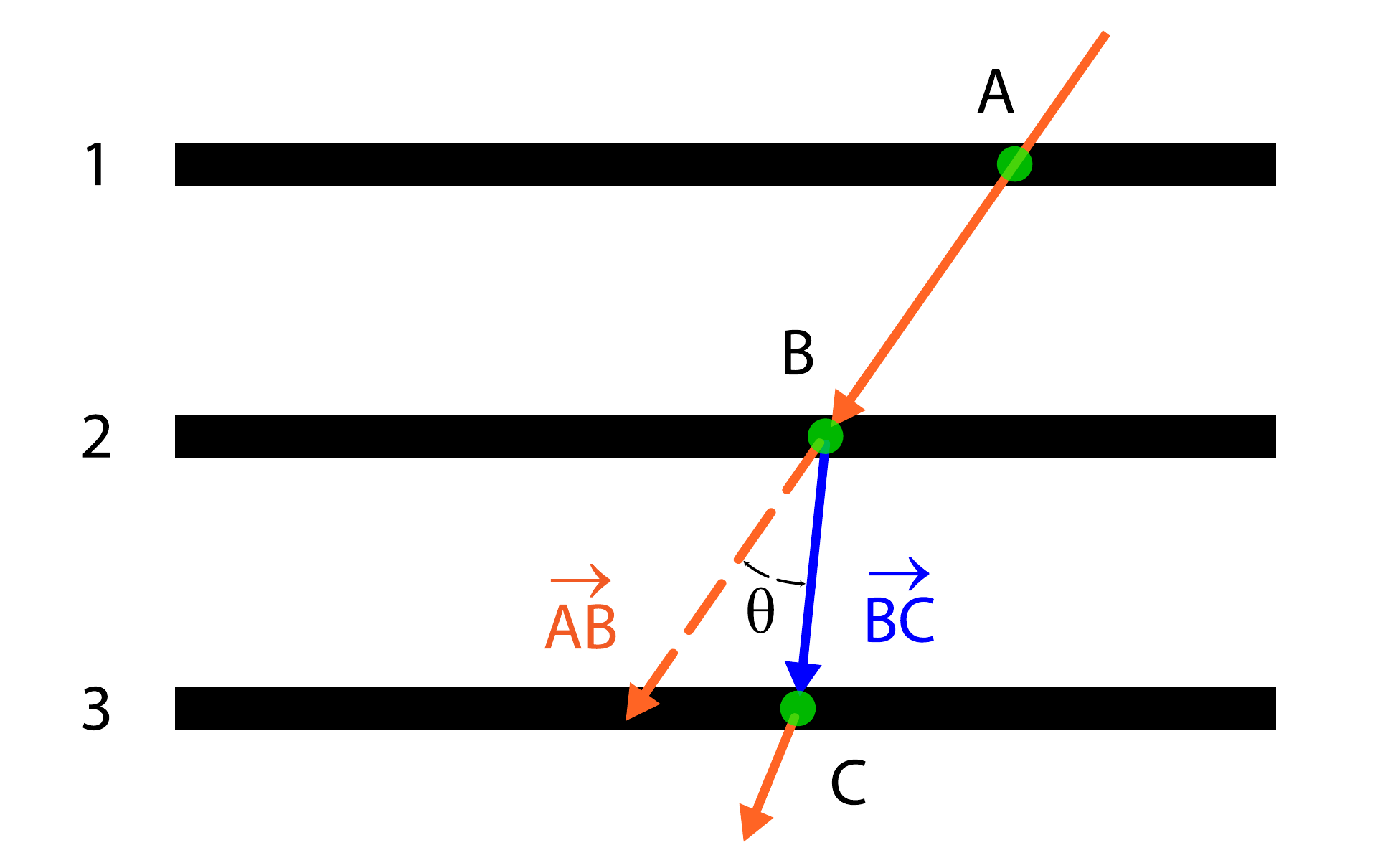}
\caption{The intrinsic scattering angle $\theta$ we use to classify the particle events (passing muons/electrons) in the hodoscope. The hodoscope has the three detector plates (black bold lines numbered as 1, 2, 3). The arrows denote the reconstructed particle trajectory. The angle $\theta$ denotes the intrinsic scattering angle of the particle in the plate 2.}\label{Fig:2}
\end{figure}

The hodoscope has three detector plates in order to estimate the energy and the type of the passing particle using the intrinsic scattering angle $\theta$ (in the hodoscope). Each detector plate has a well-defined thickness and material thus the probability distribution of the intrinsic scattering can be calculated. Though the typical $\theta$ of passing muons is in the order of mrad it can be measured in the hodoscope. As a result we can classify the reconstructed particle events (with some statistical probability) into separated categories according to their scattering angle $\theta$ in the hodoscope. Fig.~\ref{Fig:2} illustrates the intrinsic scattering $\theta$ in the hodoscope.

\begin{figure}[ht]
\centering\includegraphics[width=1.0\linewidth]{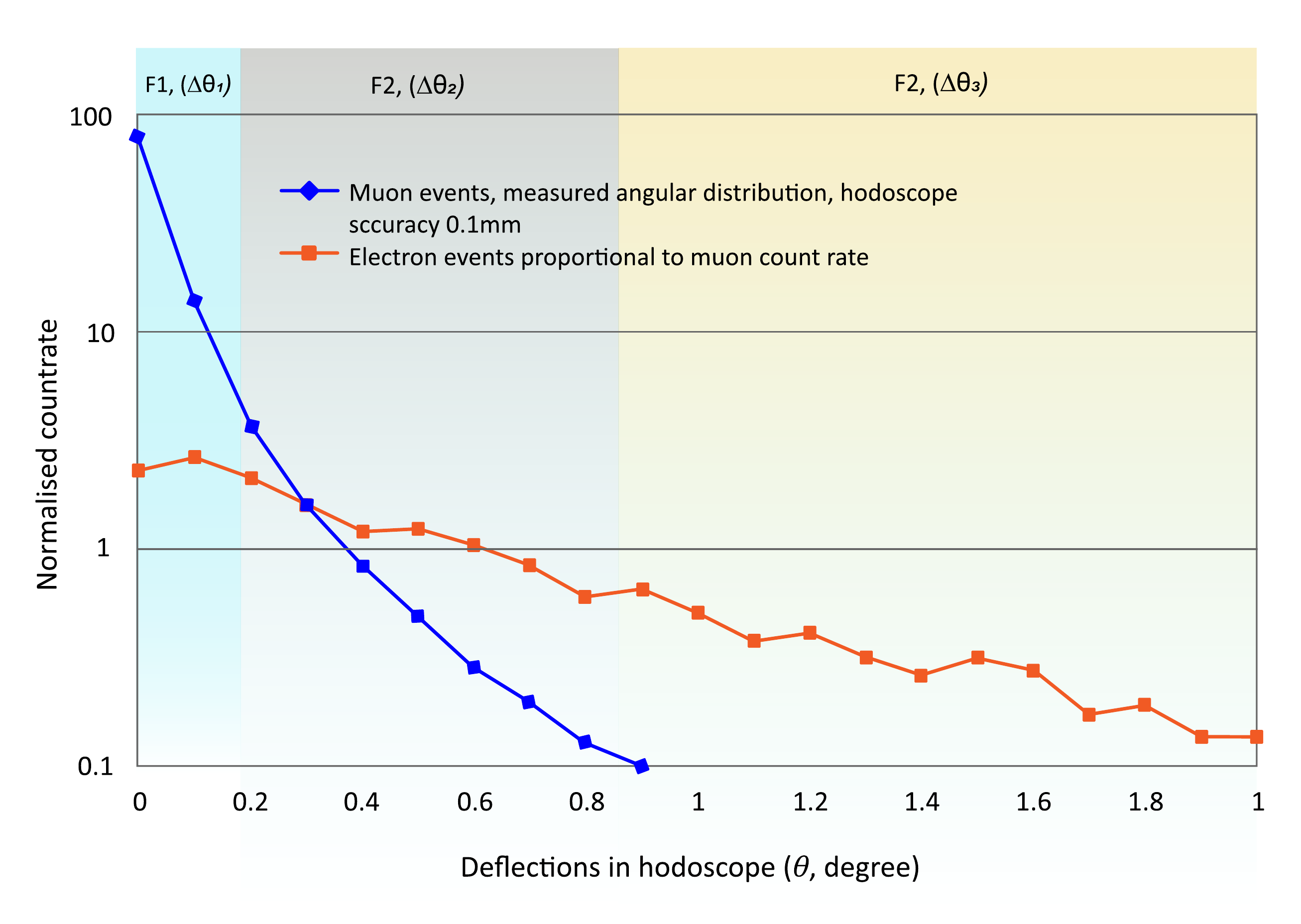}
\caption{The distribution of atmospheric ray muons and electrons as a function of the intrinsic scattering angle $\theta$ in the hodoscope (from the Geant4 model with the CRY event generator). The distribution shows we can apply the intrinsic scattering angle $\theta$ as a discriminating parameter classifying the type and energy range of the hodoscope passing particle. The latter improves the tomographic reconstruction of scanned samples very significantly. The colored areas denote the muon and electron dominated values of $\theta$ (blue, yellow) and the mixed region (gray).}\label{Fig:3}
\end{figure}

Fig.~\ref{Fig:3} shows the simulation results of the filtering spectrum for the hodoscope having the distance 100~mm between the two adjacent plates. We used the CRY cosmic ray event generator to model the atmospheric ray flux consisting of muons and electrons at sea level~\cite{CRY-paper:2007aaa}. We fixed the spatial resolution of detector plates at 0.1~mm, which corresponds to the angular resolution of 1 mrad for particles approaching the hodoscope orthogonally. Considering the angular resolution of the hodoscope, the total spectral range presented in Fig.~\ref{Fig:3} can be divided into different number of groups; we call those PTF groups below. For example, a possible robust PTF classification schema is to classify the particle events to muons and electrons or the muons with low, medium and high momentum. In Fig.~\ref{Fig:3} we have separated the spectrum into the three PTF groups: F1 (dominated by muons), F2 (mixed muons and electrons) and F3 (dominated by electrons). This classification schema has been used in the figures of the reconstruction results in the next sections.

\section{Tomographic reconstruction: the MMTR method}
\label{Sec:3}

\subsection{Reconstructing the volume of interest}

\begin{figure}[ht]
\centering\includegraphics[width=1.0\linewidth]{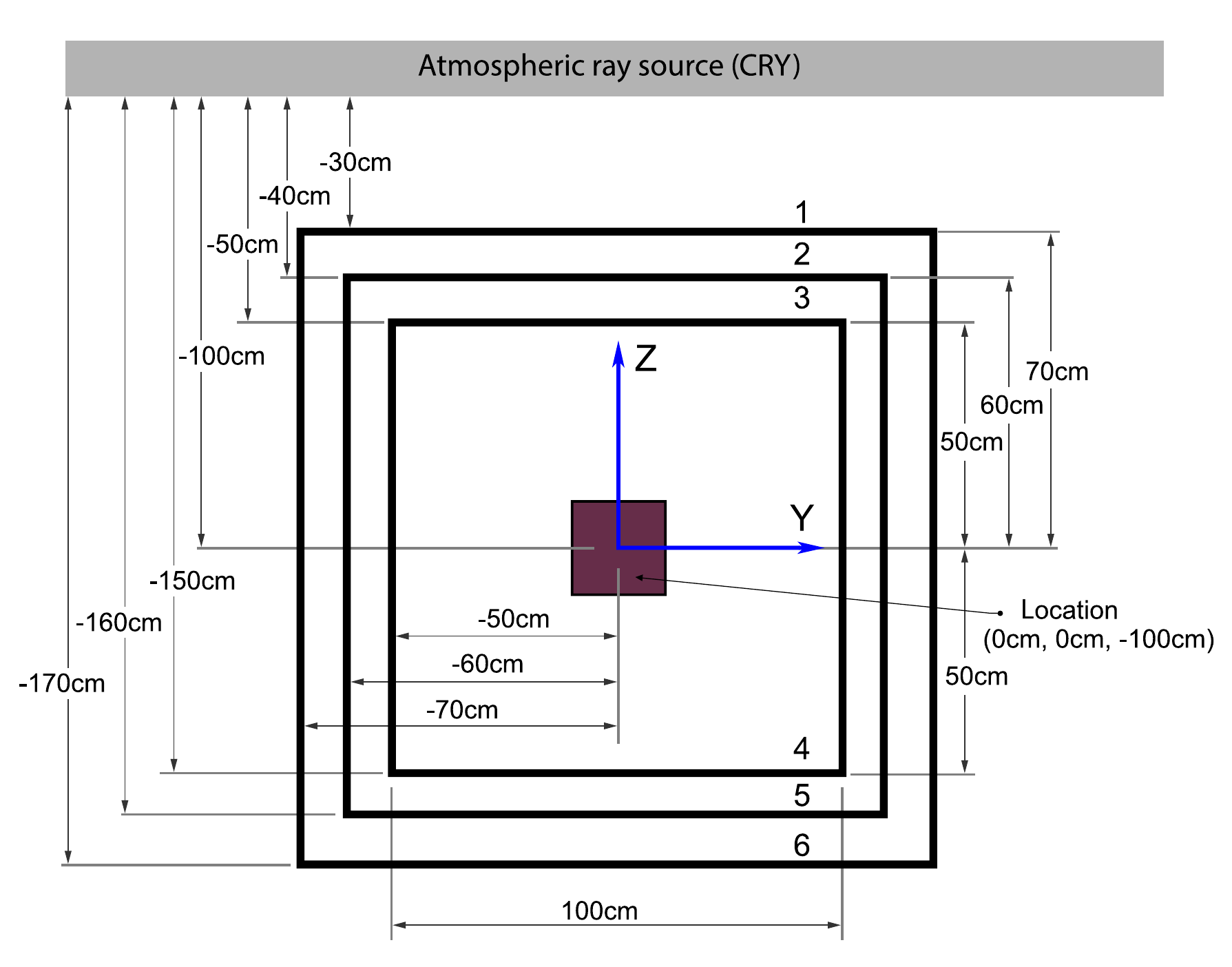}
\caption{A schema of the Geant4 model of the tomography system with the four hodoscopes (upper and lower horizontal and left and right vertical bold dark lines), the radiation source (CRY, gray area above) and an example of studied objects in the center of the system (dark purple). The numbers 1-6 denote the horizontal detector plates.}\label{Fig:4}
\end{figure}

With the design of the hodoscopes as described above, we developed a model of the tomography system in Geant4 with the interrogation volume $1 \times 1 \times 2$~m$^3$. The motivation of the size comes from practical applications, e.g., it would be suitable for small luggage, postal parcels or human body. Fig.~\ref{Fig:4} shows the dimensions of the simulated tomography system. The system consists of the two horizontal and the two vertical hodoscopes forming an enclosed interrogation volume aka the volume of interest (VOI). When a muon/electron passes the entrance hodoscope it provides the entrance incidence angle, the intrinsic scattering angle for PTF and the entrance trajectory to the VOI with its entrance location. In the same way one can characterize an exiting particle from VOI.

The geometry of the tomography system maximises the usage of side flux: on the $y$-axis the side hodoscopes measure the low incidence angle flux and on the $x$-axis the elongated hodoscopes with the 2~m length increase the field of view in the central part of VOI.

In our Geant4 simulations the exposure time was estimated by the internal code of CRY, which calculates the elapsed time on the generation of selected number of particles (all types) per the defined generation area, $10 \times 10$~m$^2$ in our case. The tomography system was centred below the CRY source. The CRY generation area was placed 30~cm from the upper hodoscope.

\begin{figure}[ht]
\centering\includegraphics[width=1.0\linewidth]{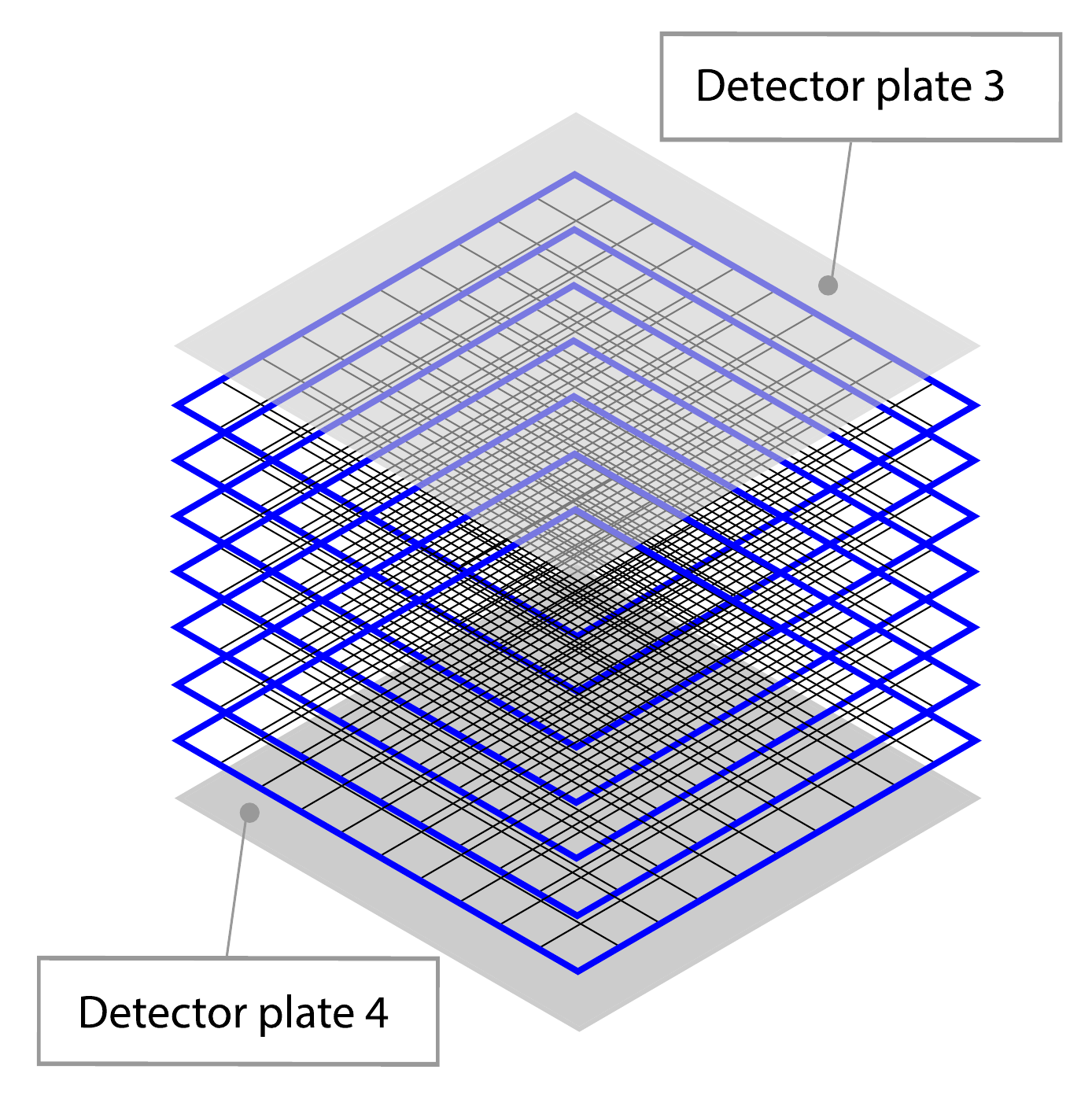}
\caption{The $z$-planes (blue) filling the volume of interest (VOI) and the voxel grid (black). The number 3 refers to the closest upper detector plate to VOI and 4 refers to the closest lower detector plate to VOI (see the numbering of the plates in Fig.~\ref{Fig:4}).}\label{Fig:5}
\end{figure}

\begin{figure}[ht]
\centering\includegraphics[width=1.0\linewidth]{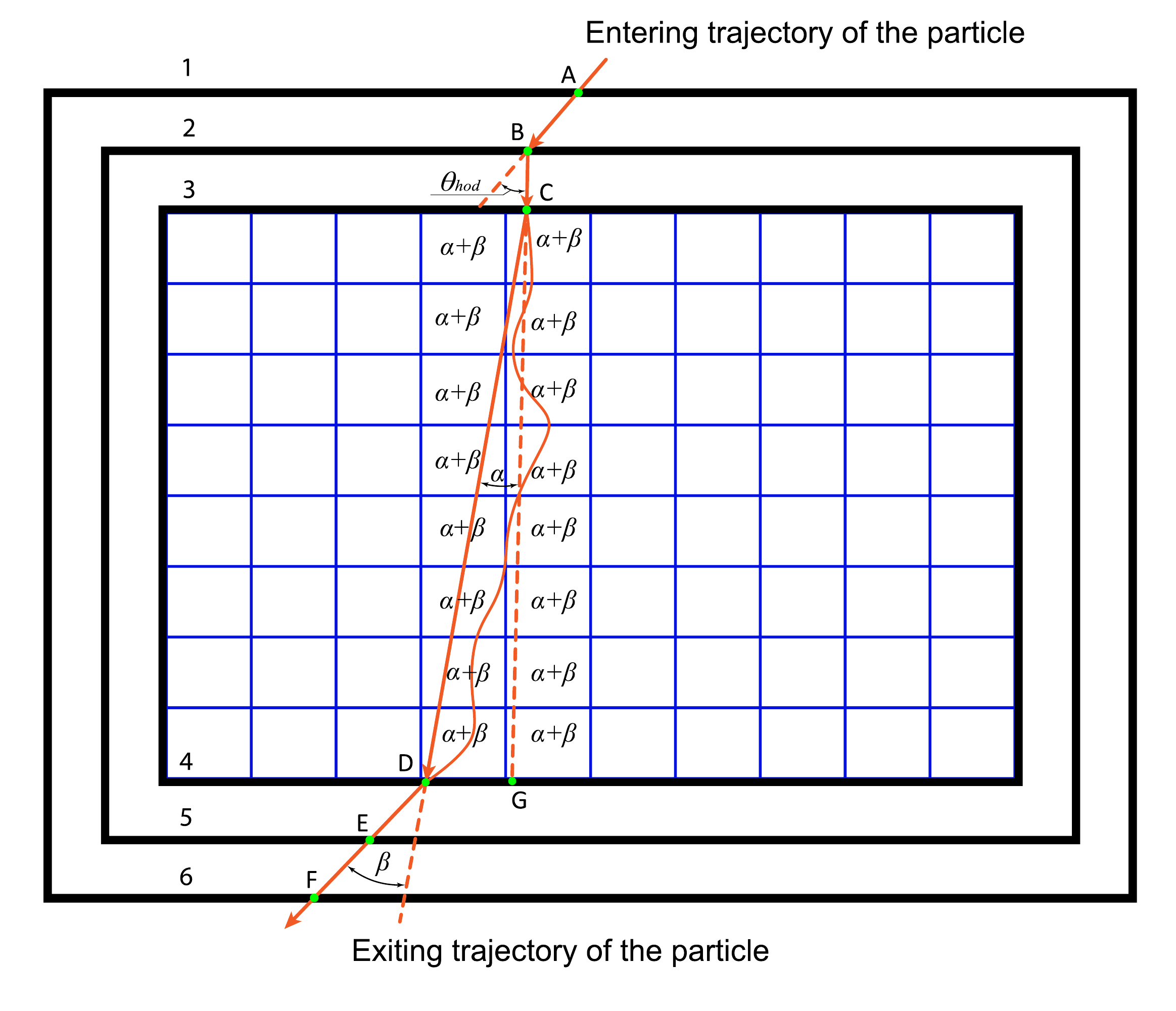}
\caption{The real and reconstructed particle trajectories through VOI in the Geant4 simulated tomography system. The designating scores to the voxels that the reconstructed trajectory passes through shown.}\label{Fig:6}
\end{figure}

In order to test PTF, we implement a tomographic reconstruction procedure based on the back projection. Thus we fill VOI with a voxel grid. In the back-projection, we use the $z$-planes as a reference point and calculate the corresponding $x$- and $y$-coordinates of the particle on a particular $z$-plane. This process is repeated for all $z$-planes, up to filling the VOI, see Fig.~\ref{Fig:5}. The $z$-plane is formed from the voxel grid point at the $z$-coordinate. The amount of the $z$-planes depends on the voxel size set by the user. As we focus on low-$Z$ materials and moderate volumes we approximate the trajectories of the particles with a straight line between the entering and the exiting locations of a particle (at the boundary of VOI). Then we allocate so called score value to each voxel when a trajectory passes through the voxel (see Fig.~\ref{Fig:6}).

We can apply the back-projection procedure to the transmission and the scattering parameters. In the case of transmission based reconstruction, the count of particle trajectories in the every voxel can be used as the score value. In the case of scattering based reconstruction then one can choose the score value as the total scattering angle or the scattering angle per unit length of the trajectory. The total scattering angle is defined as the angle between vectors BC and DE, see Fig.~\ref{Fig:6}. Alternatively, other back-projection parameters are possible (see Sec.~\ref{Sec:5}).

Summing up all particle trajectories one forms the volume density map (VDM) of VOI. We build many VDMs, a dedicated VDM per PTF group. In the case of scattering parameter, we calculate the median or the mean scattering angle for each voxel forming the VDM in the scattering based reconstruction.

It is relevant that one has different options how to combine the detector hit data in the different plate detectors. Not all the reconstructed trajectory candidates are complete, i.e. with hits in every plate detector of the entrance and the exiting hodoscope. One can have different strategies how to deal with incomplete trajectories. For simplicity, we remove all the candidates that have been registered only in the entering hodoscope. If a candidate has hits in every plate of the entering hodoscope and at least a hit in a plate of the exiting hodoscope (detector plates 1, 2, 3 and 4 or 5 or 6, see Fig.~\ref{Fig:6}) it forms so-called incomplete trajectory candidate. To apply the scattering based backprojection, at least the two hits in the plates are needed in the entrance and the exiting hodoscopes. We consider the complete and incomplete candidates in the reconstruction.

After the considering trajectory candidates we apply PTF as described in Sec.~\ref{Sec:2}. We remind that the PTF procedure classifies the candidates into three or more PTF groups. After that we feed the groups to the optimised tomographic reconstruction procedure per group. For example, we can apply different back-projection parameters on different groups.

The third step of the tomographic reconstruction is the composing of VDM. To reduce the amount of possible false trajectories we define the upper and lower limit values for the total scattering angle. Thus so far our reconstruction results the number of VDMs depending on the set of the PTF classification groups, back projection parameters etc.

\begin{figure}[ht]
\centering\includegraphics[width=1.0\linewidth]{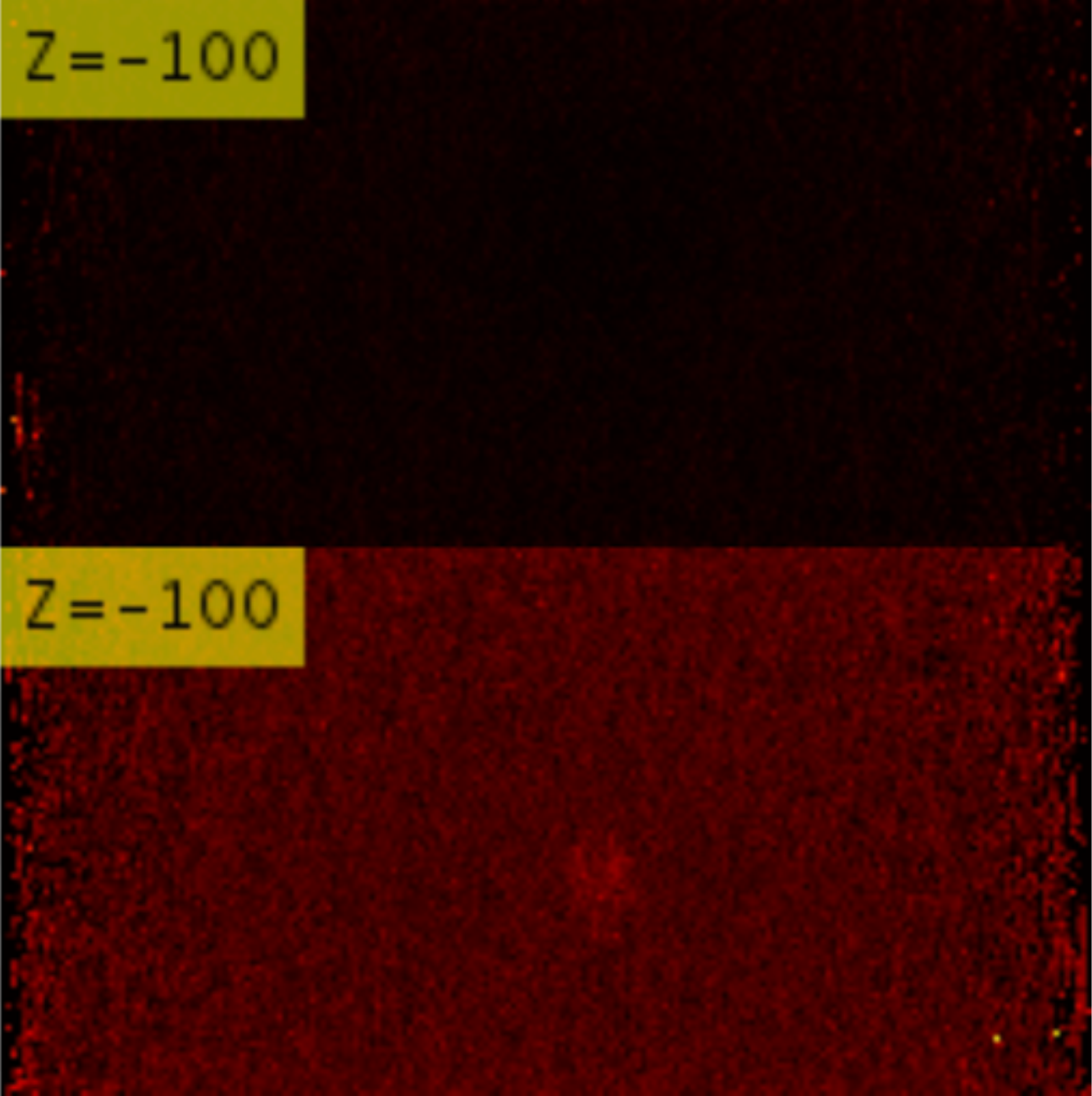}
\caption{The effect of hodoscope and VOI filtering. A cross-sectional Geant4 snapshots on horizontal plane ($z = -100$) of a cube of water ($10\times10\times10$~cm$^3$) without (upper) and with (lower) PTF. The hodoscope filtering range $0 < \theta < 2$~mrad.}\label{Fig:7}
\end{figure}

\begin{figure}[ht]
\centering\includegraphics[width=1.0\linewidth]{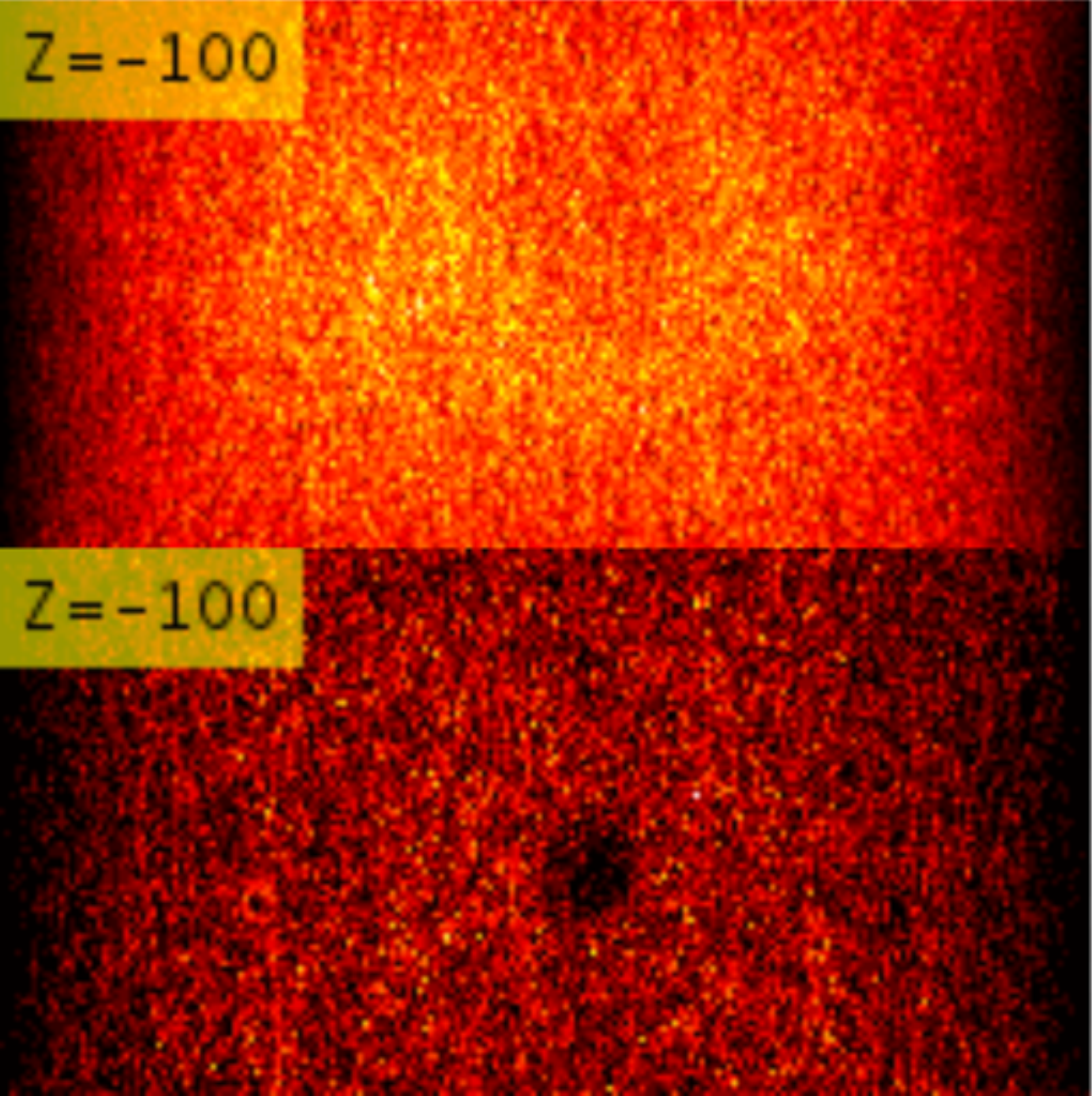}
\caption{The effect of hodoscope and VOI filtering. A cross-sectional Geant4 snapshots on horizontal plane ($z = -100$) of a cube of water ($10\times10\times10$~cm$^3$) without (upper) and with (lower) PTF. In the lower panel, the hodoscope filtering range is $10 < \theta < 30$~mrad.}\label{Fig:8}
\end{figure}

Fig.~\ref{Fig:7} and \ref{Fig:8} shows the effect of filtering parameters on the composing of VDM. The figures compare the two reconstructions of VOI (VDMs) with and without the hodoscope and VOI filtering using scattering and transmission parameter, accordingly. The figures show the VDMs at z=100~cm, where the plane crosses the interrogated object at the half height. The object is a cube of water with the side length of 10~cm centered in the middle of VOI. The estimated real measurement time is approximately 18 minutes, the voxel size $1 \times 1 \times 1$~cm$^3$. The upper panels present the reconstruction procedure with no filtering. The lower panel of Fig.~\ref{Fig:7} presents the reconstruction, where the PTF extracts the highest momentum fraction (F1, see Fig.~\ref{Fig:3}) dominated by muons. The lower panel of Fig.~\ref{Fig:8} shows the reconstruction results, where the PTF extracts the lowest momentum fraction (F3, see Fig.~\ref{Fig:3}) of muons and the most of the electrons. We obtained both the figures from the same Geant4 simulation.

The figures demonstrate that despite the reduced flux in the lower panels the filtering can significantly increase the contrast of VDMs. In those examples we have separated the two physical effects: the absorption and scattering. Thus the different settings of the filtering parameters and ranges can potentially increase the discrimination capability for object detection and material classification.

\subsection{Object detection}

In this subsection we describe the algorithms we developed to extract object shapes from VDMs. A crucial property of the algorithms is that they have to operate well in high Poisson noise conditions. We also notice that due do proprietary information the methodology of the edge detection procedure is disclosed with some generality.

\begin{figure}[ht]
\centering\includegraphics[width=1.0\linewidth]{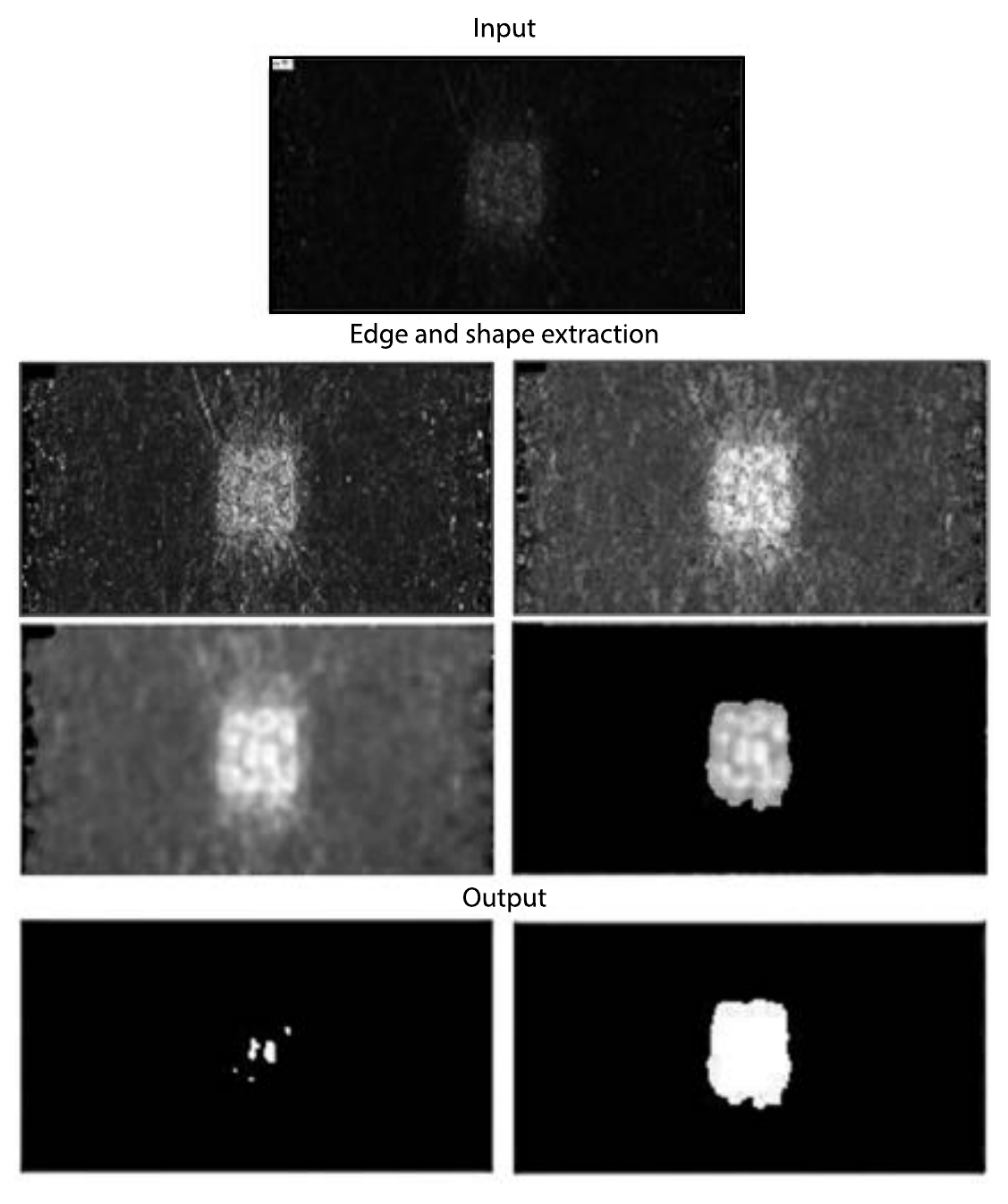}
\caption{Example of the double-object detection algorithm. From top to down, the original reconstructed image down to the detected objects through the iterative object detection procedure in case of double-object system: a cube ($10\times10\times10$~cm$^3$) of RDX explosive material surrounded by a larger cube of flesh ($30\times30\times30$~cm$^3$). The estimated measurement time in the Geant simulation is 18~minutes. The voxel size is $1\times1\times1$~cm$^3$, the spatial resolution of the detector plates 0.1~mm.}\label{Fig:9}
\end{figure}

We start the object reconstruction on the 2D data. An algorithm enhances the edges of objects from VDM iteratively. After several iterations, the image sharpness is enhanced to a level, where we apply an image threshold to extract the binary image, the final reconstructed shape of the object. Fig.~\ref{Fig:9} shows a double-object example, a smaller inner cube of RDX-explosive with the side length of 10~cm surrounded with a larger cube of flesh-like material with the side length 30~cm. As described above we apply PTF. In this case the VDM represents the distribution of mean scattering angles. The spatial resolution of the plates in the Geant4 simulation was 0.1~mm. The estimated measurement time in the Geant simulation is 18~minutes.

\begin{figure}[ht]
\centering\includegraphics[width=1.0\linewidth]{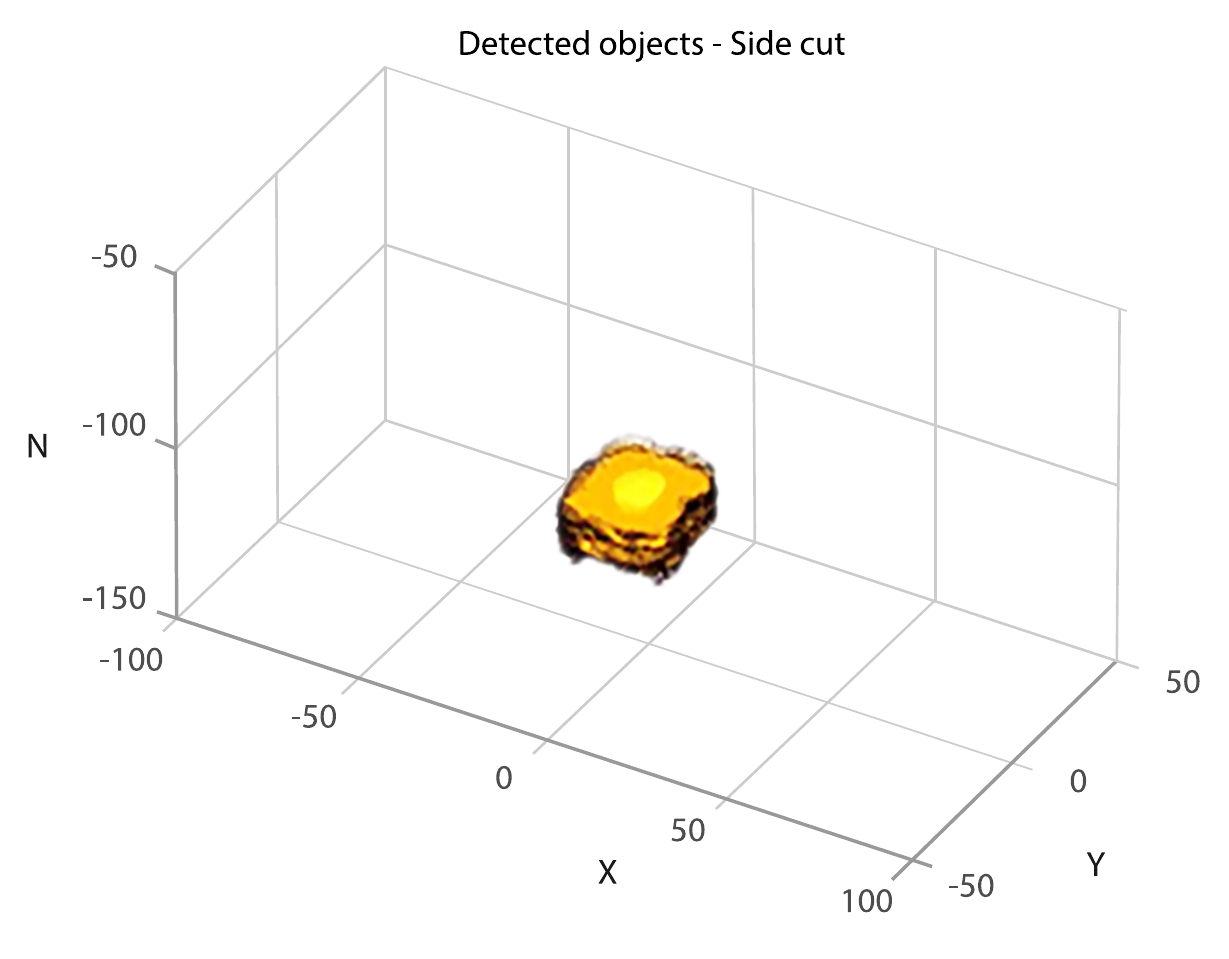}
\caption{The object reconstruction and detection of a double object with spectroscopic filtering relying on the scattering density parameter. The double object: a cube of RDX explosive material ($10\times10\times10$~cm$^3$) centered in a cube of flesh ($30\times30\times30$~cm$^3$), the voxel size is $1\times1\times1$~cm$^3$. The object detection and separation to the two logical volumes are denoted by yellowish colors. The spatial resolution of the detector plate is 0.1~mm.}\label{Fig:10}
\end{figure}

After the 2D reconstruction, the results are fed to the 3D reconstruction procedure. To differentiate between the object candidates, we introduce a labelling step that gives a unique index to the connected volumes. This can be achieved having so-called Connected-Component Labelling (CCL), where the subsets of connected components are uniquely labelled using, e.g., Two Pass Algorithm [33]. Fig.~\ref{Fig:10} presents a half-cut of reconstructed and detected objects (for the same simulation as in Fig.~\ref{Fig:9}). The result indicates clearly that the VDMs of the two materials are distinguishable enough for the edge detection and the two different logical volumes have been established. The two different logical volumes are denoted: yellow for RDX, orange and black edges for flesh.

\begin{figure}[ht]
\centering\includegraphics[width=1.0\linewidth]{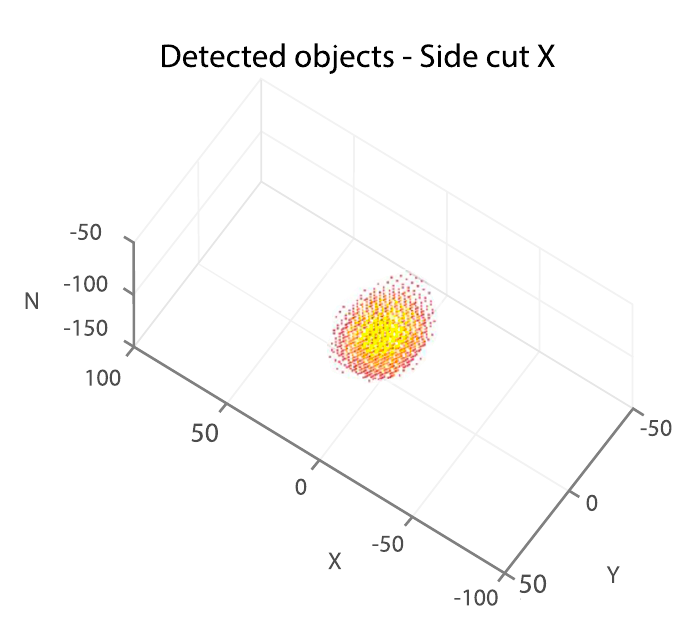}
\caption{The object reconstruction and detection of a double object with spectroscopic filtering relying on the scattering density parameter. The double object: a cube of RDX explosive material ($10\times10\times10$~cm$^3$) centered in a cube of flesh ($30\times30\times30$~cm$^3$), the voxel size is $3\times3\times3$~cm$^3$. The object detection and separation to the two logical volumes are denoted by yellowish colors. The estimated measurement time 2 minutes and the spatial resolution of the detector plate 0.1~mm.}\label{Fig:11}
\end{figure}

We tested the same object configuration with a shorter, 2-minutes estimated, measurement time. Fig.~\ref{Fig:11} shows the results. One can clearly see the loss of contrast. Due to lower statistics, we increased the voxel size from 1~cm$^3$ to 9~cm$^3$. However, the edge detection algorithm is still able to detect and separate the two materials as different logical volumes.

Once we establish and separate the logical volumes of different object we can apply special algorithms for the material or other classifications on the reconstructed volumes. Those will be the subject of future work.

\section{Design of the physical prototype}
\label{Sec:4}

\subsection{Technical description}

\begin{figure}[ht]
\centering\includegraphics[width=1.0\linewidth]{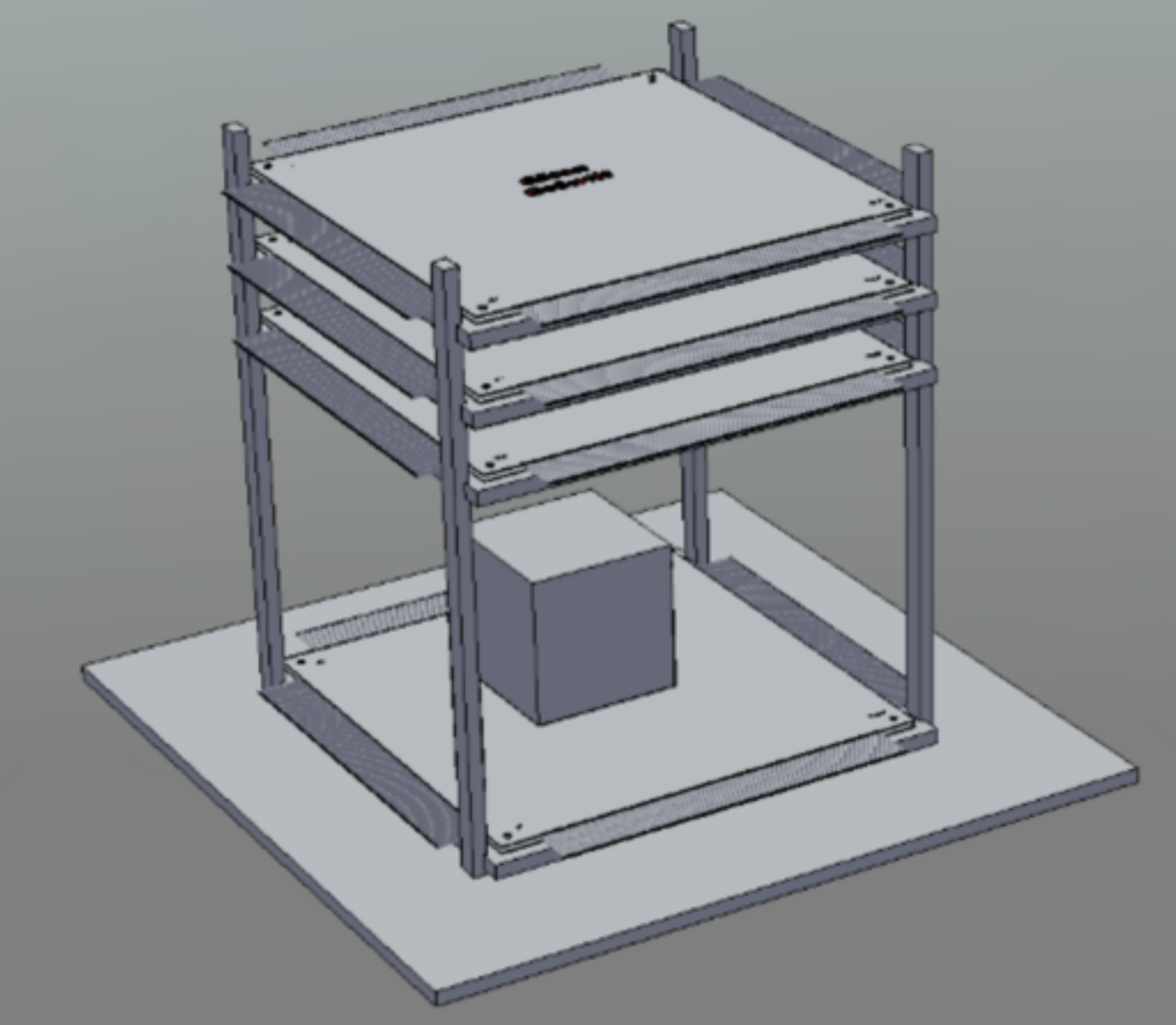}
\caption{The general spatial design with the detector plates and the structural elements. An example object (gray cubic) is in the middle of the detector between the upper hodoscope (the three horizontal plates) and the lower single detector plate (a plate below the cube). The fiber-mat couplings with SiPMs nor the DAQ-system are not shown.}\label{Fig:12}
\end{figure}

As described above we developed the PTF and MMTR methods based on the virtual Geant4 models. To prove (and improve) the methods we built a simplified physical prototype system. Fig.~\ref{Fig:12} shows the principal design of the physical prototype. The prototype has an upper hodoscope with three detector plates based on the above described fiber-mat technology. The lower hodoscope has been replaced with a single detector plate. This design means the system cannot measure the total scattering angle. However, one can estimate the total scattering angle by calculating the angle between the vectors BC and CD, see Fig.~\ref{Fig:6}. This angle depends on the position of the object in a vertical axis, thus one can do a comparison material classification experiment fixing objects into the same position in every measurement. An another shortcoming of the minimal design is that the detector plates cover only a small fraction of the total solid angle around VOI and only a limited fraction of the flux is available. The detector plate size is $25 \times 25$~cm$^2$, the distance between the two adjacent plated in the upper hodoscope is 7.5~cm. The volume of VOI is $25 \times 25 \times 25$~cm$^3$. Thus the field of view is limited to 64~degrees in $x$- and $y$-axis. The limited field of view results in low vertical reconstruction accuracy if no correction algorithm implemented.

First, we studied the detection and reconstruction performance of the prototype in the Geant4 simulations. We describe the main detection performance parameters, detection efficiency and spatial resolution, in the following subsections.

\subsection{Particle tracking in the fiber-mat system}

\begin{figure}[ht]
\centering\includegraphics[width=0.6\linewidth]{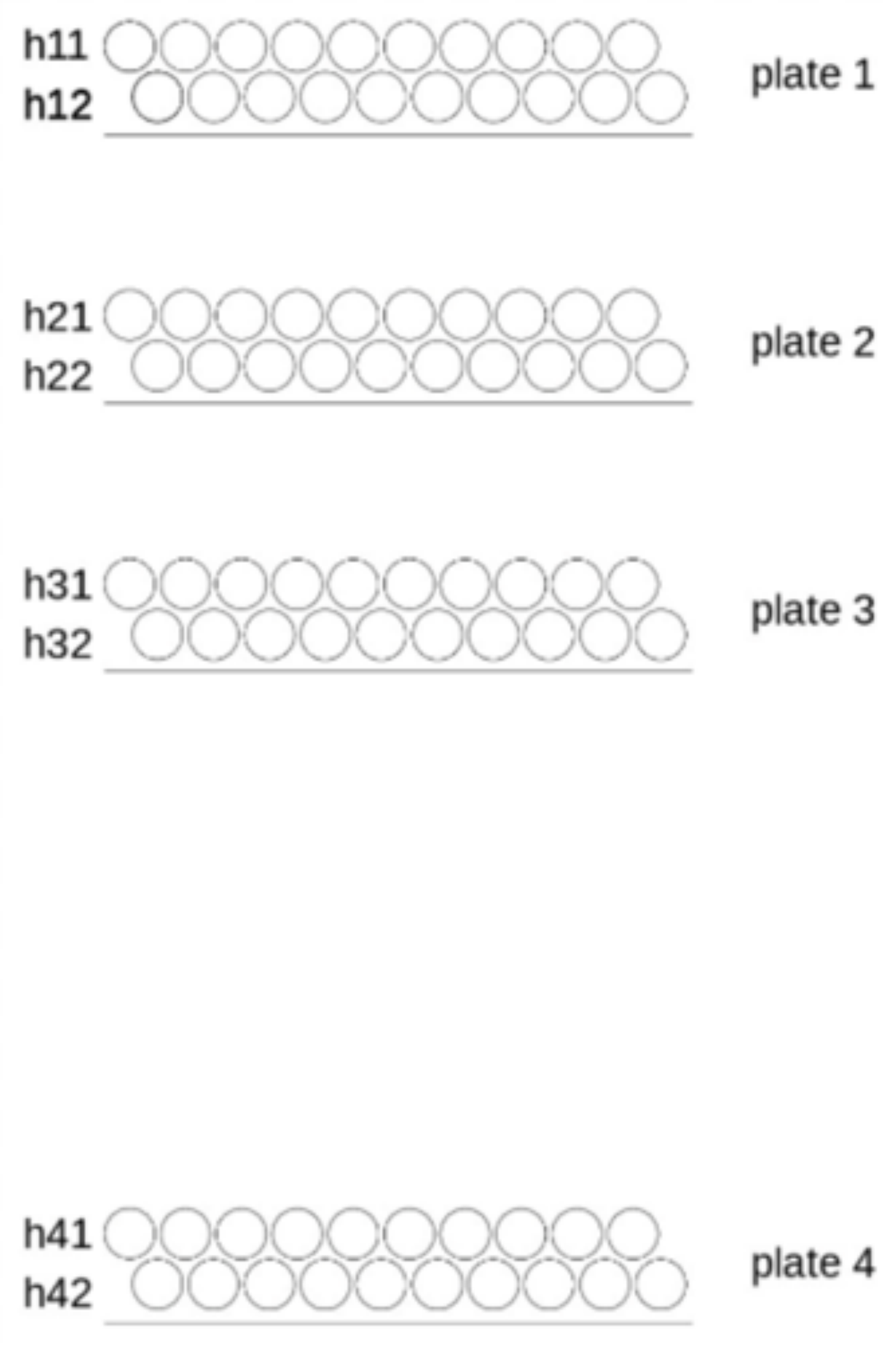}
\caption{The labeling schema of the bi-layered scintillating fibers of the detector (showing the $y$-directional by-layer only).}\label{Fig:13}
\end{figure}

In this section we explain how we build up a particle track candidate form hit candidates (lit fibers). First, we explain our notations and the coordinate system. For simplicity, we deal with the $x$- and $y$-directions separately and so below we show only a direction if sufficient. Fig.~\ref{Fig:13} shows the fiber schematics and the notations through the four plates of the prototype in a selected direction ($x$ or $y$). As described in Sec.~\ref{Sec:2} and shown in Fig.~\ref{Fig:1} a plate has the four fiber layers in total arranged to the two orthogonal bi-layers. We address the fiber layers as h$AB$ where $A$ denotes the plate and $B$ denotes the layer in the plate.

For example, h11 denotes the plate 1, the y-directional bi-layer and the first layer. h14 denotes the plate 1, the x-directional bi-layer and the second layer. For the z-coordinate of the hit candidate we define a virtual plane between the bi-layers. So the layers h11 and h12 are above the virtual $z$-plane and the layers h13 and h14 are below the $z$-plane. We define the component of $x$ or $y$ layers as the fiber-centre offsets from the $z$-plane. The following parameters describe the fiber: the core diameter, the width and the pitch of the cladding. The pitch is the distance between the adjacent fiber centres. As we know all the positions and other parameters of the fibers we can compute the centres of the fibers in the well-defined 'absolute' Cartesian coordinate space. We enumerate the fibers starting from 0 to $N$ in a layer, where $N$ is the max number of fibers in the layer. Thus detecting a fiber lit we can then convert it the 'absolute' coordinates.

\begin{figure}[ht]
\centering\includegraphics[width=0.75\linewidth]{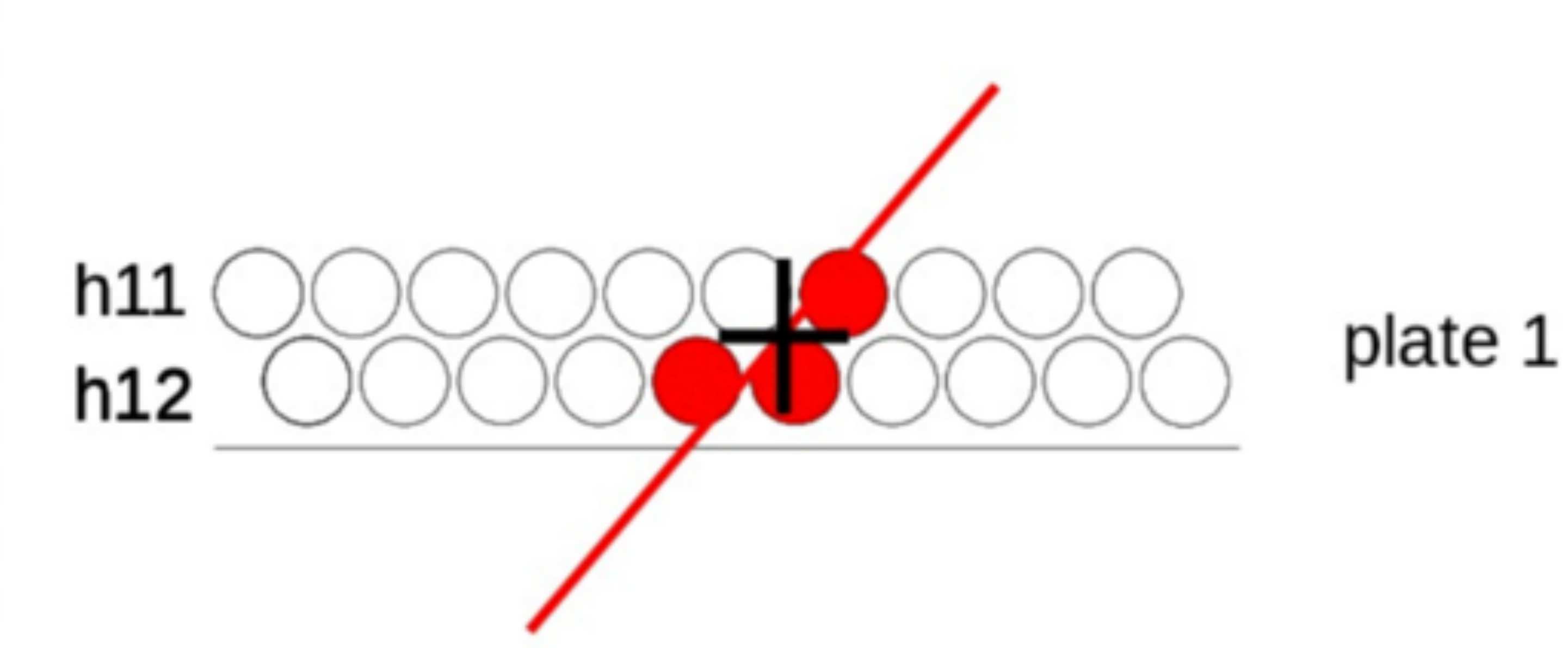}
\caption{The reconstruction of the hit candidate for the $y$-directional by-layer. The red denotes the lit fibers. The cross depicts the reconstructed hit candidate. The red line is an example of particle track candidate reconstructed having some hit candidates from the other plates.}\label{Fig:14}
\end{figure}

To reconstruct a plate hit we handle the $x$- and $y$-directions separately. For example, Fig.~\ref{Fig:14} shows the reconstructed hit for the $y$-direction from the three lit fibers in the $y$ by-layer. We compute the hit coordinate finding the weighed arithmetic mean of fiber centres lit. In the figure, red denotes the fibers lit and the black cross is the hit coordinate reconstructed. The same calculation is performed for the $x$ by-layer. The red line is an example of particle track candidate reconstructed having some other hit candidates from the other plates.

\begin{figure}[ht]
\centering\includegraphics[width=0.75\linewidth]{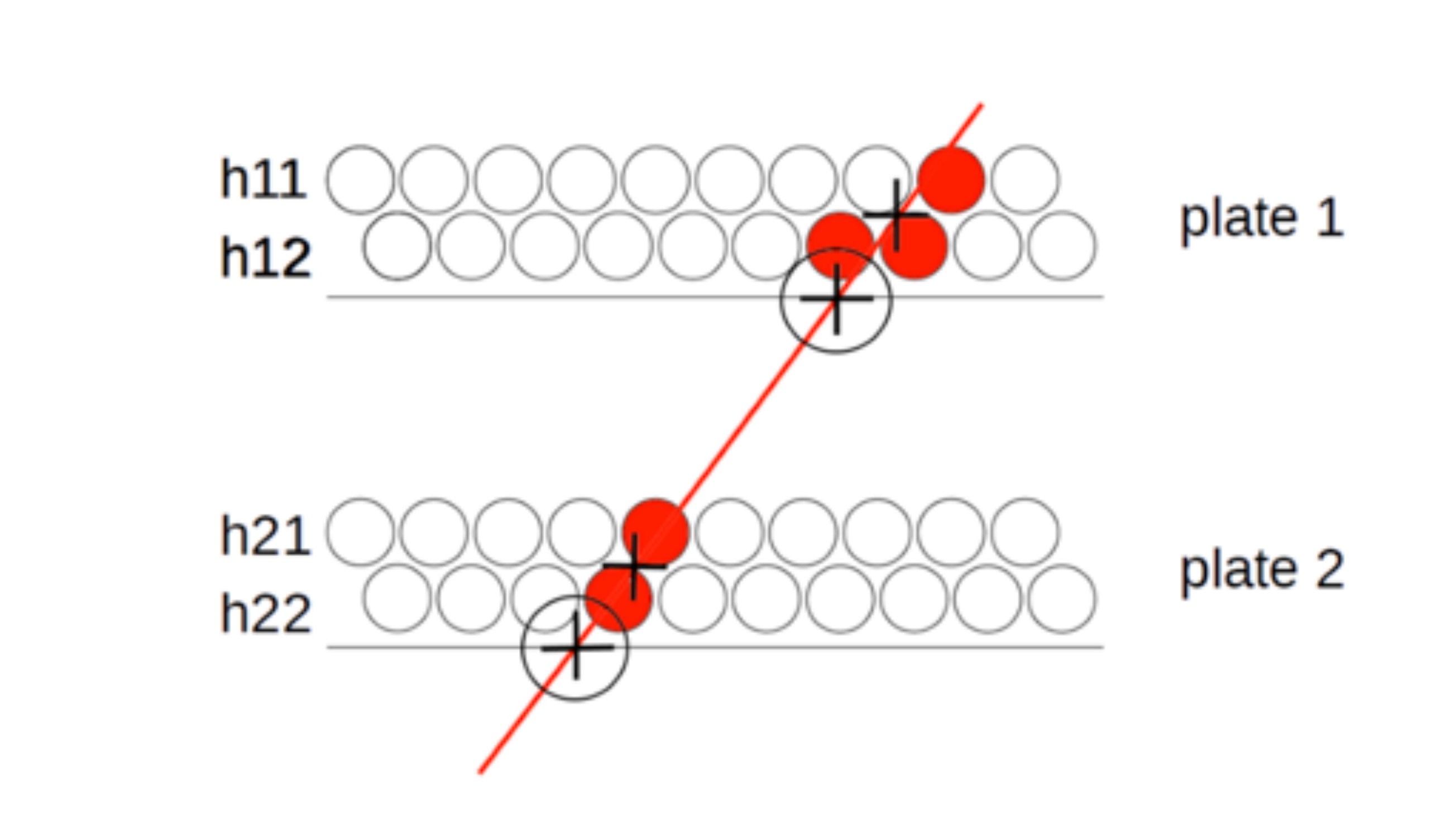}
\caption{The hit extrapolation to the plate-z. The red fibers represent fibers generating signal, the cross depicts the calculated particle hit coordinate, the circled cross indicates the extrapolated coordinate.}\label{Fig:15}
\end{figure}

The $x$ and $y$ coordinates of a hit reconstructed (at least) in the two planes we can project the results to the $z$-plane. We use the linear extrapolation based on the reconstructed track candidate. Fig.~\ref{Fig:15} shows the extrapolation results (black circled crosses). Combining the component values from the both axes projected to same $z$ plane, we have reconstructed the 3-dimensional coordinate of a hit candidate.

\begin{figure}[ht]
\centering\includegraphics[width=0.75\linewidth]{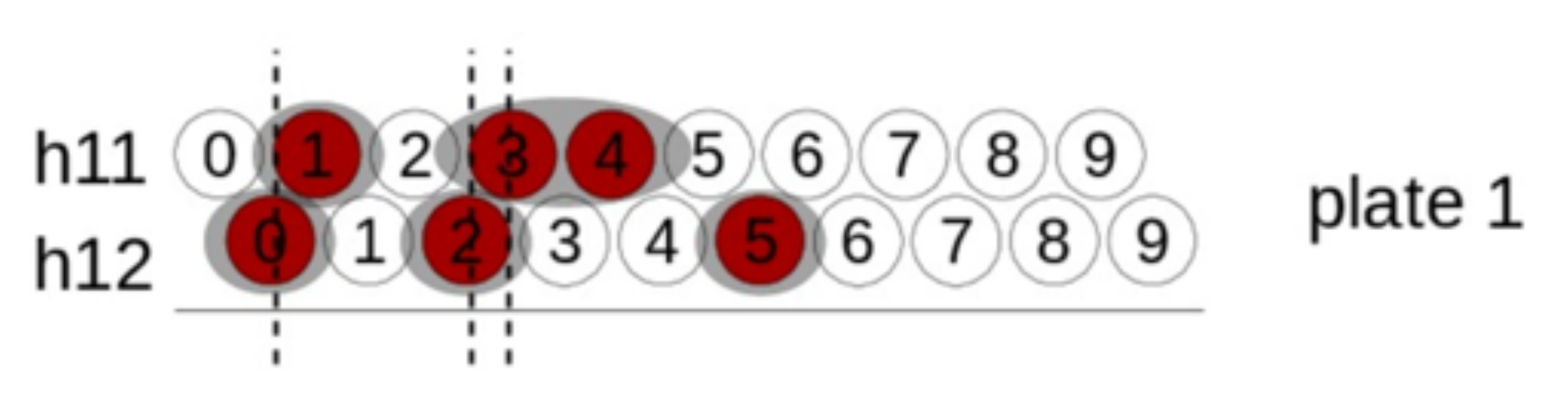}
\caption{An clustering example, the red fibers represent the fibers generating signal.}\label{Fig:16}
\end{figure}

For the track reconstruction we need a clustering algorithm to cluster the lit fibers from the weighted mean of fiber centres computed. Fig.~\ref{Fig:16} shows some cases of clustered lit fibers to illustrate the approach. Clustering algorithm works in the two stages. First, we cluster the lit fibers from the single layers. We form clusters of consecutively lit fibers. In the layer h11 the fibers are {1, 3, 4} and in the layer h12 the fibers {0, 2, 5}. Second, we merge the clusters from the different layers if they intersect component-wise: either the $x$ or $y$ coordinates depending on the axis are handled. Fig.~\ref{Fig:16} shows this as the dotted intersecting region. We compute the minimum extreme of a cluster by taking the leftmost fiber, computing its intersections in the fiber layer $z$-plane and taking the minimum value. The same logic is followed to calculate the maximum extreme value. 

\subsection{Geant4 simulations for the prototype}

We tested the hit and track reconstruction algorithms described in the last sub-section in the Geant4 simulations of the prototype. To estimate the quality and effectiveness of the algorithms we collected the following information of the muon/electron hits in the virtual detector plates: (i) the hit coordinates and (ii) the number of fiber(s) providing the signal from the same event for the every detector plate and layer. We generated the simulation data with a statistically sufficient number of events having the CRY-based source described above. Then we calculated the distance differences between the Geant4 provided coordinate values (i) and the coordinates provided by our algorithms calculated for every $x$ and $y$ layer and all the simulated events. We calculated the distributions and the standard deviations, some results are presented in Table~\ref{Tab:1}. The numbers in the table are based on the 13 699 simulated hit events: 10 907 muon events and 2792 electron events. The distribution has its mean at zero and it corresponds to the Laplace distribution.

\begin{table}[h!]
 \begin{tabular}{c|c c c} 
 &
 \multicolumn{3}{c}{\textbf{Stdev} ($\mu$m)} \\
 \cline{2-4}
 \textbf{Detector plate} & \textbf{Muons} & \textbf{Electrons} & \textbf{All} \\ 
 \hline 
 Plate 1 & 113 & 1234 & 567 \\
 Plate 2 & 110 & 726 & 343 \\
 Plate 3 & 153 & 840 & 1112 \\
 Plate 4 & 109 & 2769 & 1261 \\
 \hline
 \textbf{Average} & \textbf{121} & \textbf{1392} & \textbf{821} \\ 
\end{tabular}
\caption{Standard deviations between the hit locations and reconstructed hit locations from the Geant4 simulations of the prototype tomography system.}
\label{Tab:1}
\end{table}

In the case of muons, the standard deviation varies slightly above 100~$\mu$m. The reason of this rather large value is in a limited set of extreme cases, which have 1-2 mm deviations from the actual hit coordinates. In the case of electrons, we expect larger deviations due to more scattering events of electrons in the detector plates, decreasing the performance of the track reconstruction algorithm. However, the number of strongly deflecting extreme event is considerably small. If we remove some tens of most deflecting events, the standard deviation falls down to a few hundreds of $\mu$m.

\begin{figure}[ht]
\centering\includegraphics[width=1.0\linewidth]{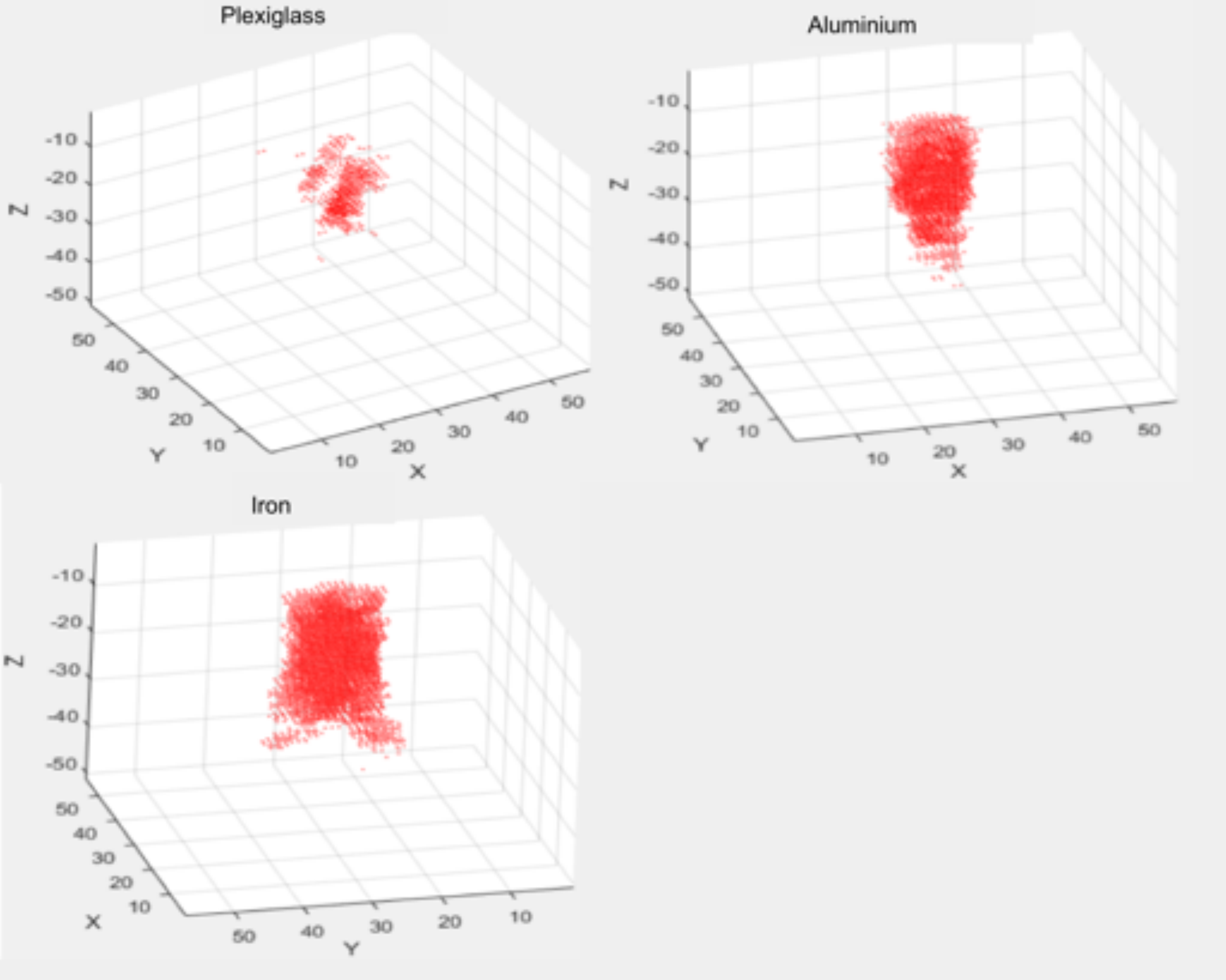}
\caption{The reconstructed logical volumes by the prototype tomography system. The scanned object (plexiglass, aluminium, iron) has the size $6\times6\times6$~cm$^3$. The voxel size $0.5\times0.5\times0.5$~cm$^3$ and the estimated measurement time 30~minutes. The scale of the coordinates is denoted in voxels (e.g. 10 corresponds to 5~cm).}\label{Fig:17}
\end{figure}

\begin{figure}[ht]
\centering\includegraphics[width=1.0\linewidth]{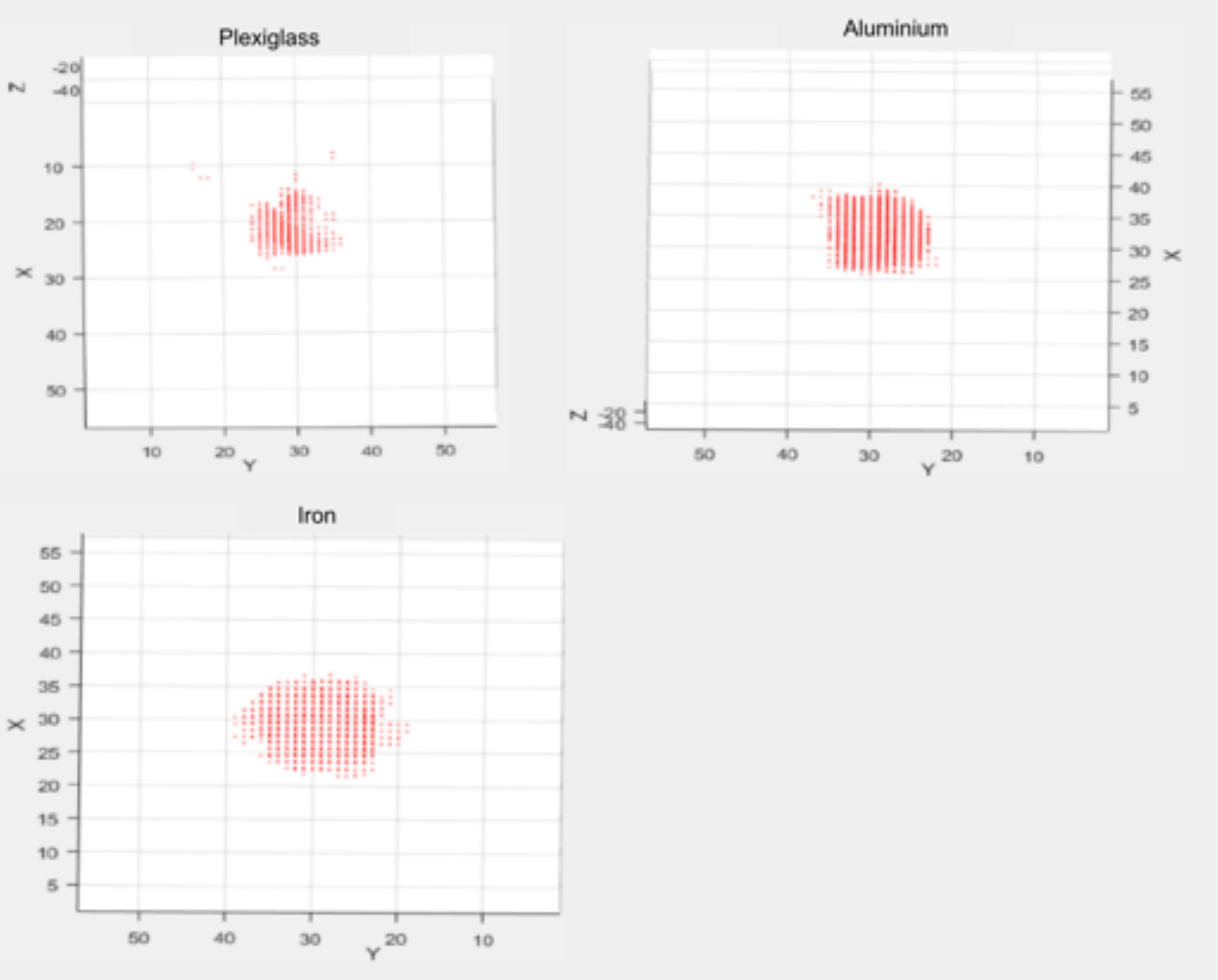}
\caption{The horizontal cross-sections of reconstructed logical volumes at the half-height of the object applying (the same reconstruction schema as in Fig.~\ref{Fig:17}). The scanned object (plexiglass, aluminium, iron) has the size $6\times6\times6$~cm$^3$. The voxel size $0.5\times0.5\times0.5$~cm$^3$ and the estimated measurement time 30~minutes. The scale of the coordinates is expressed in voxels (e.g. 10 corresponds to 5~cm).}\label{Fig:18}
\end{figure}

To test the performance of the tomographic reconstruction in the Geant4 simulations we had different cube shaped materials in VOI. We fixed the location of the object in the same position size and had the $6 \times 6 \times 6$~cm$^3$ cube shaped objects. We centered the object in the horizontal plane and located the objects vertically 20~mm from the lower detector plate of the upper hodoscope. Fig.~\ref{Fig:17} and \ref{Fig:18} show the reconstruction results of the three different materials. To create the VDMs we applied the mean scattering angle. The estimated real measurement time was 30~minutes.

The horizontal cross-sections have been reconstructed quite accurately, for lead object it is overestimated about 10~mm, for Plexiglas 10~mm less than its actual size. Due to the limited angle tomography, reconstructed objects are vertically elongated about 2-3 times their actual size.

\section{The physical prototype and the first measurements}
\label{Sec:5}

\subsection{Construction of the prototype}

\begin{figure}[ht]
\centering\includegraphics[width=1.0\linewidth]{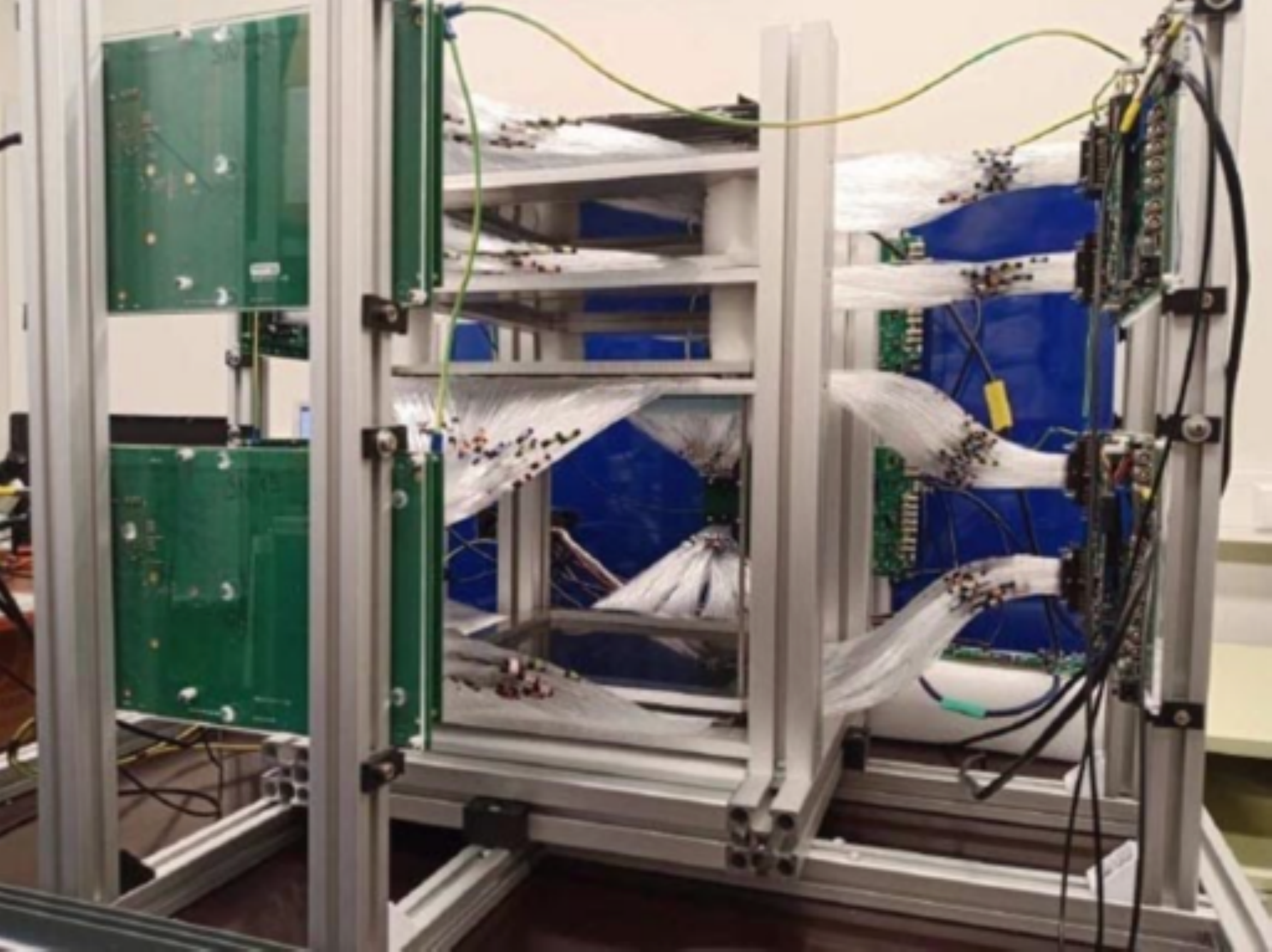}
\caption{The prototype tomography system is encased into a light-tight box. The upper casing plate removed for the photograph.}\label{Fig:19}
\end{figure}

Based on the design of the prototype tested with Geant4, we constructed the physical prototype to have a physical validation of the proposed detector systems, PTF and MMTR. As in the case of virtual model, the physical prototype (see Fig.~\ref{Fig:19}) has the upper hodoscope with three detector plates and the lower hodoscope with single detector plate. We have the detector layers with the 1~mm round single cladding scintillation fibers (Saint-Gobain BCF-12) assembled as shown in Fig.~\ref{Fig:1}. The active area of each detector plate is $247 \times 247$~mm$^2$. We spaced the detector plates in the upper hodoscope with 75~mm pitch, the distance between lower plate of the upper hodoscope and the lower hodoscope is 250~mm and thus the dimensions of the VOI are $247 \times 247 \times 250$ mm$^3$.

We installed the detector plates into a rigid inner frame that fixed the aluminium support frames of the detector plates. The external frame is used to fix the positions of the SiPM arrays and the DAQ boards. On top of the upper hodoscope and below the lower hodoscope, we assembled the two trigger detectors (scintillator plates).

\begin{figure}[ht]
\centering\includegraphics[width=1.0\linewidth]{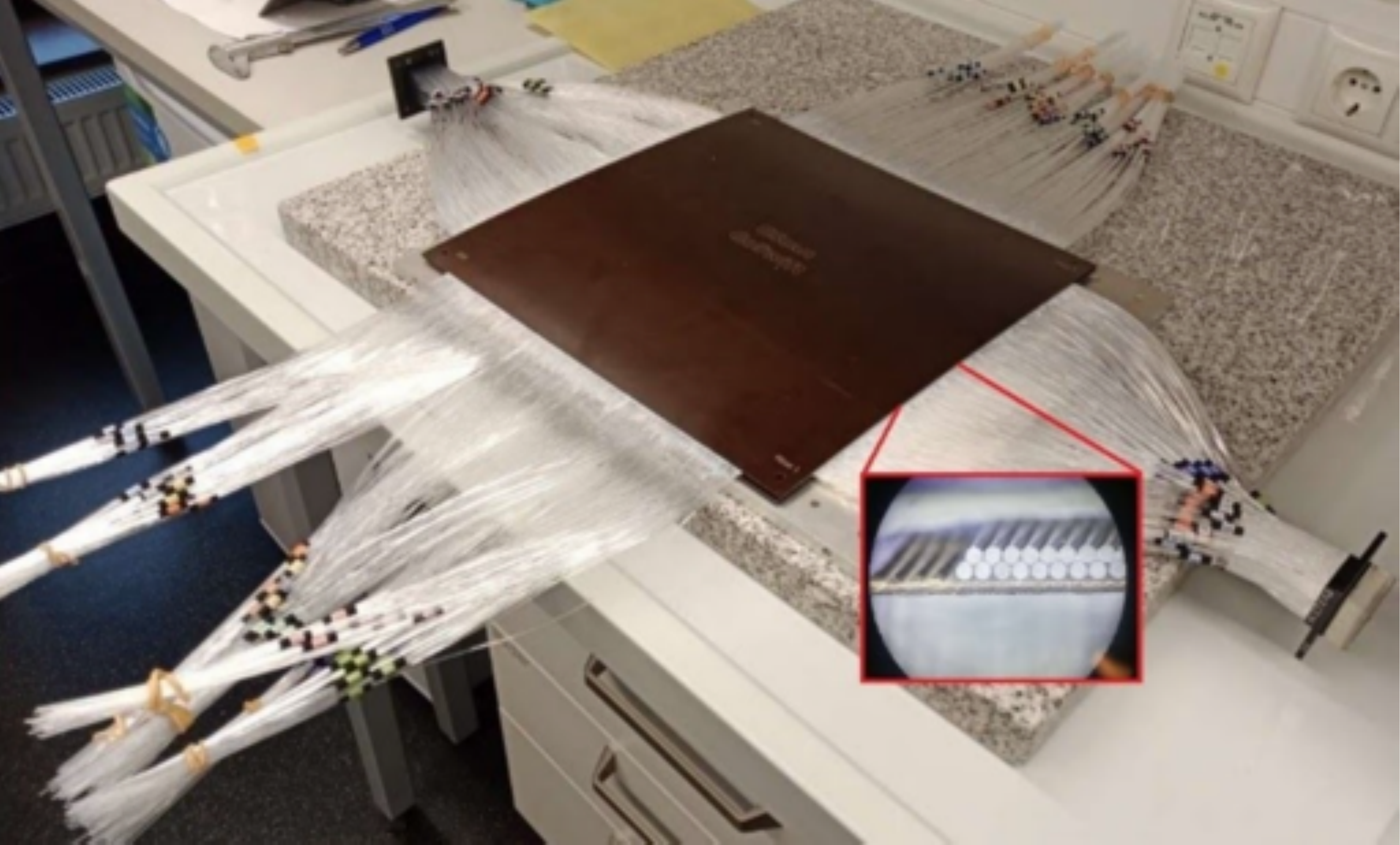}
\caption{The assembly of a detector plate. The inset shows the magnified front image of the grooved surface of the support plate and the fiber positions in the fiber mat.}\label{Fig:20}
\end{figure}

We assembled the detector plates using grooved alignment tables (see Fig.~\ref{Fig:20}). This guarantees the accurate and controlled positioning of fibers in the fiber mat. The alignment tables have been produced from 1~mm thick sheets of phenolic paper laminate (PFCP201) by milling the grooves with a pitch of 1.100~mm. We glued the fibers on the alignment table using UV-curing glue (Loctite AA 3311). We added TiO2 powder to the glue to reduce optical leakage between the fibers.

We assembled the two separately produced double-layered fiber mats orthogonally to each other in order to form a 2D-sensitive four-layered detector plate. We supported the assembly with an aluminium frame that provides sufficient mechanical rigidity, flatness and fixation with the frame. At the corners of each PFCP201 plate we have the alignment holes allow to align the two orthogonal 2D fiber mats against each other and to align the four detector plates with respect to each. For the latter we used the alignment rods once the detector plates were mounted in the frame. The structure ensures a mechanical tolerance for relative plate positions below $\leq0.1$~mm. We estimate the mechanical tolerance of the fiber positioning on the fiber mat to be below 0.01~mm and the orthogonality of the two fiber mats below 0.001~degrees.

\subsection{Data Acquisition System}

We composed the data acquisition system of the prototype implementing eight CAEN DT5550W boards. We use Ketek (PA3325-WB-0808) and Hamamatsu (S13361-3050AE-08) SiPMs arrays to collect scintillation light from the fibres.

Two plastic scintillator plates trigger the data readout. The active area of a trigger is $30 \times 30$ cm$^2$, the upper one placed on top of the upper hodoscope and the second one placed under the lower hodoscope. The triggers have two $6 \times 6$~mm$^2$ SiPMs (Hamamatsu S13360-6050CS) using the SiPM Readout Front-End Board CAEN A1702. When a muon/electron passes the trigger system (both plastic scintillator plates) a trigger signal is created that initiates the readout of all the CAEN DT5550W boards.

\subsection{Data analysis: testing PTF and MMTR}

Having the physical prototype our first objective was to test the detection efficiency of the completed particle tracks. The efficiency is defined as a completed particle track candidate reconstructed if the triggered signals are collected from all the 8 bi-layer fiber mats. The average count rate from the triggering system is about 3~Hz indoor (the lab room has concrete walls and ceilings and one additional floor above). Considering the triggering detector area (0.09~m$^2$) and the field of view (66 degrees) of the double trigger plate telescope, the average count rate is in the expected range. We performed a 30-minute measurement with the empty VOI volume to estimate the efficiency of the completed tracks. In the test we recorded the 6594 triggered events, out of which the 3562 were completed tracks. It means the total efficiency for the completed particle tracks is around 50\%.

\begin{figure}[ht]
\centering\includegraphics[width=1.0\linewidth]{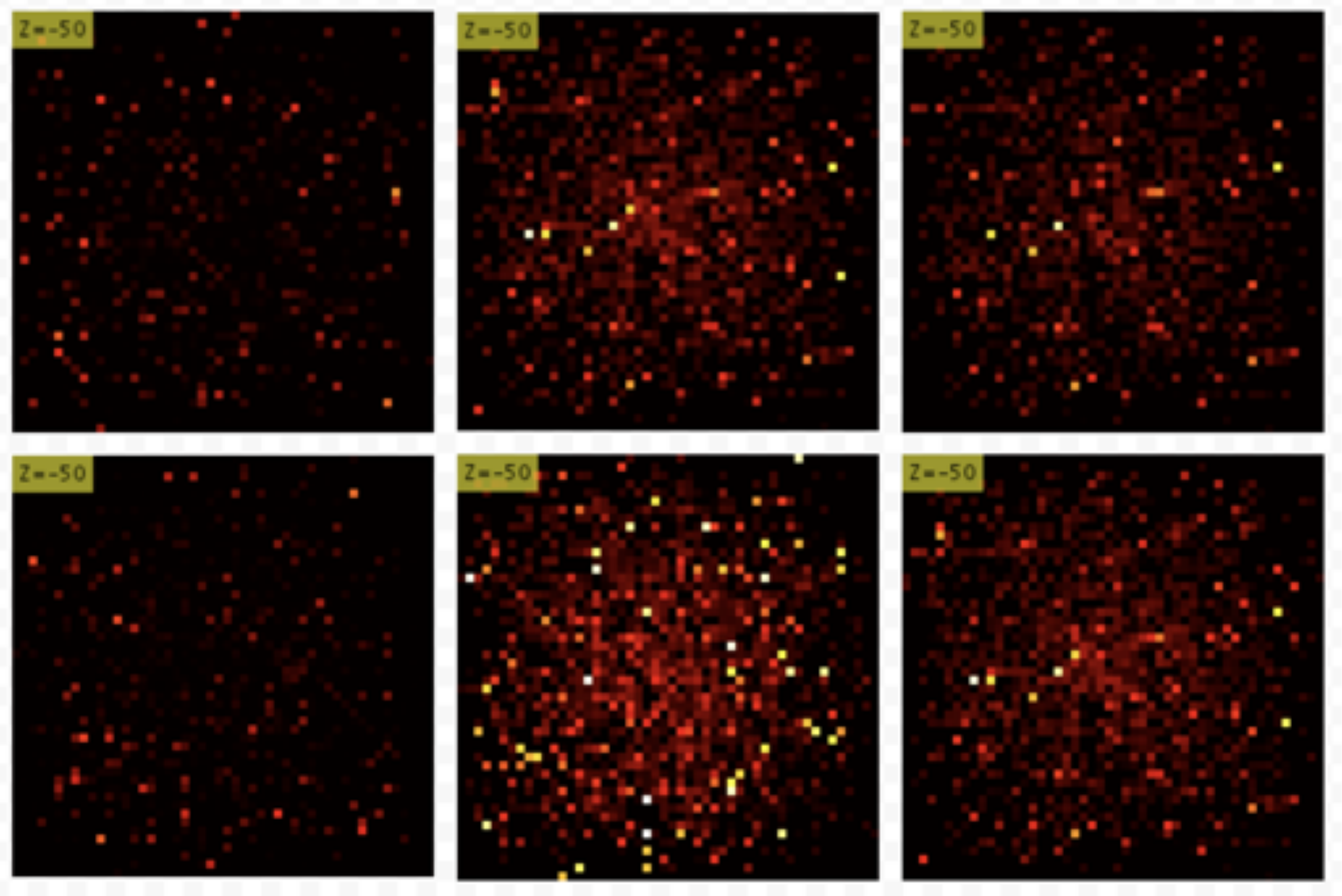}
\caption{The horizontal cross-sections with the different filtering conditions for a cube of water and aluminium ($6\times6\times6$~cm$^3$): no hodoscope nor VOI filtering applied (left); no hodoscope filtering applied (middle); hodoscope and VOI filtering applied (right).}\label{Fig:21-22}
\end{figure}

Fig.~\ref{Fig:21-22} shows the comparisons of the reconstruction having the different filtering conditions on the 2500 completed trajectories. The voxel size was $0.5 \times 0.5 \times 0.5$~cm$^3$, the reconstruction parameter was the mean scattering angle. Due to the limited field of view, the horizontal cross-sectional images are a little unclear, different from the simulation results of the full model of the tomography system. After we applied PTF, the contrast of the reconstructed images improved, especially in the case of light materials.

\begin{figure}[ht]
\centering\includegraphics[width=1.0\linewidth]{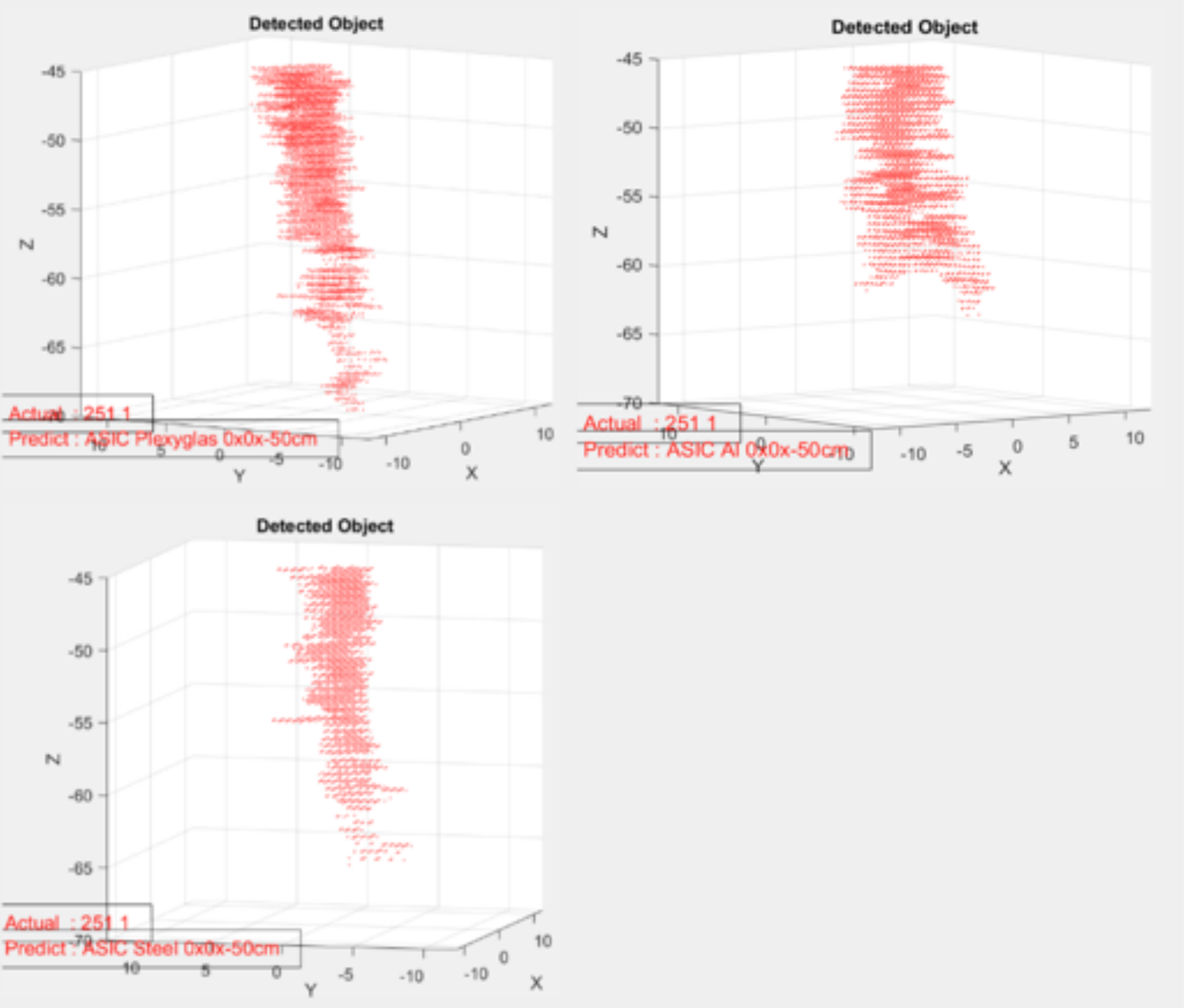}
\caption{The detected logical volumes as a result of the reconstruction of a cube samples ($6\times6\times6$~cm$^3$) of plexiglass (upper left), aluminium (upper right) and steel (lower left). The samples were centred 2~cm below the lower detector plate of the upper hodoscope. The red region represents the logical volume for the detected object. The scale of the coordinates is in cm.}\label{Fig:23}
\end{figure}

\begin{figure}[ht]
\centering\includegraphics[width=1.0\linewidth]{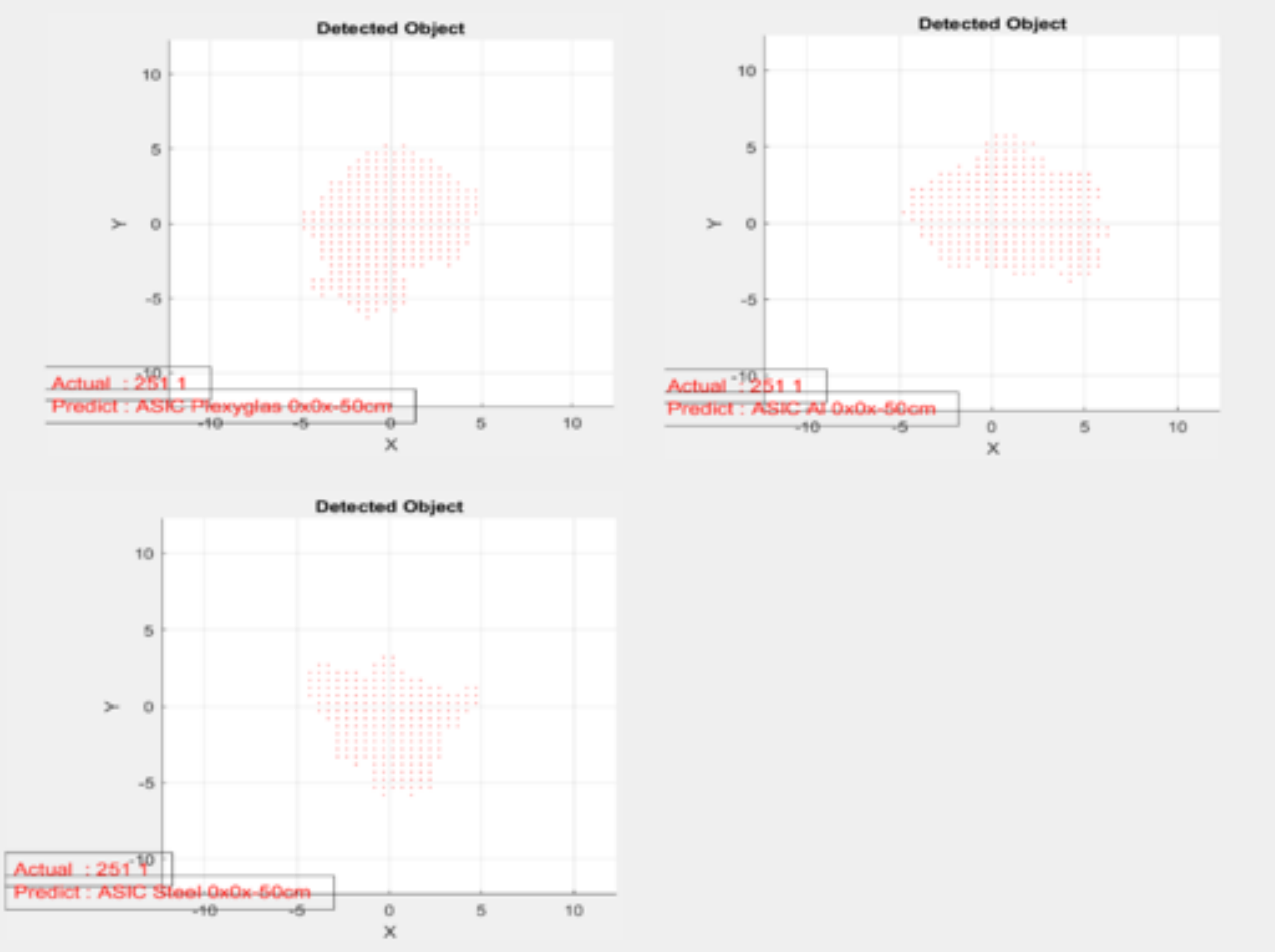}
\caption{The horizontal cross-sectional cuts of the same samples as in Figure~\ref{Fig:23}. The cut is at the half-height of the object and the reconstruction schema same as in Figure~\ref{Fig:23}. The scale of the coordinates is in cm.}\label{Fig:24}
\end{figure}

Having the optimised PTF and MMTR parameters, we reconstruct the 3D images for the three different cases: a cubic of aluminium, Plexiglas and ammonium nitrate (in granular form) with the dimensions $6 \times 6 \times 6$~cm$^3$ (see Fig.~\ref{Fig:23} and \ref{Fig:24}. The reconstruction parameter for VDMs was the mean scattering angle.

The reconstruction results are vertically elongated as we predicted from the Geant4 simulations and vary between 15-25~cm. Horizontally, the side length vary between 7-8~cm. The dimensions of the objects are overestimated, which is expected due to limited angle tomography. Theoretically, one can apply a special correction if needed.

\section{Discussions and conclusions}

In this paper, we discussed on the two new methods in muon tomography: PTF (the classification of particle candidates in the hodoscope) and MMTR (the creating of multi-VDMs following the classification schema of PTF, etc). Our main motivation is to improve object reconstruction and material classification for low-Z materials in the case of ART. We applied the methods on the hodoscope system based on plastic scintillating fibers. We proposed the detector technology to ensure the required detector performance, mainly the spatial accuracy, the sensitivity and the lightness of the detector system. First, we presented the results based on the Geant4 simulations.

Second, we validated the methods on the minimal table-top prototype tomography system. We tested the performance of PTF and MMTR on the prototype having different low-Z materials. We showed that applying the proposed methods we can successfully detect the low-Z objects in a 10–30-minute time scale. For example, we tested the objects composed by water, Plexiglas, ammonium nitrate (mimicking explosive material). Our Geant4 simulations indicate that one can reduce the exposure time to a few minutes range having the full-scale tomography system (see Sec.~\ref{Sec:3}) based on the same technology like the physical minimal prototype.

The particle tracking algorithm determining the particle hit location in a detector plate and its trajectory in the hodoscope provide the accuracy 120~$\mu$m for muons. We demonstrated that VDM can be analysed as a set of 2D images applying an edge detection algorithm on the VDM. It allows to detect the objects as separated logical volumes. The method works efficiently in the case of high Poisson noise, hence applicable for short measurement times.

When logical volumes reconstructed one can be interested in the material classification of volumes. The possible classification procedure can follow as a single procedure or having the multiple procedures, e.g., one per PTF group. For example, if one combines the classification based the filtered fractions F1 and F3 (see Fig.~\ref{Fig:1}), the discrimination ability of materials increases including both the scattering and transmission data simultaneously. Furthermore, establishing the reconstructed objects as discrete entities enables to apply material classification techniques based on machine learning.

One of our next steps is to develop the material classification algorithms on VDMs. Some potential methods to start the classification study would be machine learning based such as the Kullback-Leibler Divergence (KLD) or the Support vector Machine (SVM). Another future step will be the reconstruction of complex objects representing more realistic user cases. For example, we can implement some iterative reconstruction methods in order to increase the ability of generating VDMs for complex scanning objects.

The PTF and MMTR methods combined with the proposed detector system improve the reconstruction of VOI significantly. The combination makes the detection and potentially classification of low-Z composed materials possible in small-scale detector system. It opens up new routes for the commercialization of ART tomography. The developed methods and the detector system is interesting in the context of industrial security applications such as baggage scanning in airports and cargo and shipping container scanners for illicit materials and contraband.

\section*{Acknowledgements}

We would like to thank Andrea Giammanco for very helpful discussions and comments. This work was continuously and strongly supported by the startup company GScan OÜ. This work was supported by the Enterprise Estonia, the grant EU48693. AH thanks the Estonian Research Council for the grant PRG434 and the EC for the ERDF CoE program project TK133.




\bibliographystyle{elsarticle-num-names}
\bibliography{ART.bib}

\end{document}